\UseRawInputEncoding
\documentclass[journal]{IEEEtran} \usepackage[utf8]{inputenc} \usepackage{amsmath,amsfonts,amssymb}
\usepackage{array}
\usepackage[caption=false,font=normalsize,labelfont=sf,textfont=sf]{subfig}
\usepackage{textcomp}
\usepackage{stfloats}
\usepackage{float}
\usepackage{url}
\usepackage{enumitem}
\usepackage{verbatim}
\usepackage{graphicx}
\usepackage{cite}
\usepackage{nomencl}
\usepackage{siunitx}
\usepackage{mathtools}
\usepackage{cuted}
\usepackage{lipsum}
\usepackage[utf8]{inputenc}
\usepackage[T1]{fontenc}        % good practice for fonts
\usepackage{algorithm}
\usepackage{algpseudocode}
\usepackage{tikz}
\usepackage{threeparttable}
\usepackage{multirow}  % put this in preamble
\usepackage{makecell}
\usetikzlibrary{positioning, shapes}
\usetikzlibrary{shapes.geometric, arrows}
{\upshape} 
\newtheorem{definition}{Definition}
\newtheorem{notation}{Notation}

\newtheorem{remark}{Remark}

\makenomenclature
\nomenclature{$\rho$}{Screw lead}
\nomenclature{$\lambda$}{Screw frictional angle}
\nomenclature{$\phi$}{Screw lead angle}
\nomenclature{$n$}{Screw transmission ratio}
\nomenclature{$\eta_f$}{Screw forward efficiency}
\nomenclature{$\eta_b$}{Screw backward efficiency}
\nomenclature{$r$}{Screw radius}
\nomenclature{$\psi_{PM}$}{Magnetic flux of permanent magnet}
\nomenclature{$i_d$, $i_q$}{$d$-axis and $q$-axis currents}
\nomenclature{$v_d$, $v_q$}{$d$-axis and $q$-axis voltages}
\nomenclature{$L_d$, $L_q$}{$d$-axis and $q$-axis inductances}
\nomenclature{$R_s$}{Resistance of the stator windings}
\nomenclature{$P$}{Number of pole pairs}
\nomenclature{$\omega_m$}{Angular velocity of the rotor}
\nomenclature{$\tau_e$}{Motor electromagnetic torque}
\nomenclature{$\theta_e$}{Rotor electrical angle}
\nomenclature{$\theta_m$}{Rotor mechanical angle}
\nomenclature{$v_a$,$v_b$,$v_c$}{Three-phase voltages of motor}
\nomenclature{$k_{CR}$}{Movement conversion ratio}
\nomenclature{$f_{n}$}{Thrust force of screw}
\nomenclature{$I_{t}$}{Total inertia moment of mechanism}
\nomenclature{$V_{t}$}{Total viscous friction coefficient of mechanism}
\nomenclature{$\tau_{t}$}{Coulomb friction of motor and gearbox}
\nomenclature{$F_{t}$}{Coulomb friction of screw}
\nomenclature{$N_{g}$}{Gearbox ratio}
\nomenclature{$M_{t}$}{Total mass of screw and load}
\begin{document}
© 2025 IEEE. Personal use of this material is permitted.
Permission from IEEE must be obtained for all other uses,
including reprinting/republishing this material for advertising
or promotional purposes, collecting new collected works
for resale or redistribution to servers or lists, or reuse of
any copyrighted component of this work in other works.
This work has been submitted to the IEEE for possible
publication. Copyright may be transferred without notice,
after which this version may no longer be accessible.
\title{All-Electric Heavy-Duty Robotic Manipulator:\\ Actuator Configuration Optimization and Sensorless Control}

\author{Mohammad Bahari, Amir Hossein Barjini, Pauli Mustalahti, and Jouni Mattila
        % <-this % stops a space
\thanks{Funding for this research was provided by the Business Finland partnership project ``Future All-Electric Rough Terrain Autonomous Mobile Manipulators'' (Grant No. 2334/31/2022) and the Research Council of Finland under the Project ``Nonlinear PDE-model-based control of flexible manipulators'' (Grant No. 355664).}% <-this % stops a space
\thanks{All authors are with the Faculty of Engineering and Natural Sciences, Tampere University, 33720 Tampere, Finland. Corresponding author: Mohammad Bahari (\texttt{mohammad.bahari@tuni.fi}).}}



\maketitle
\begin{abstract}
This paper presents a unified framework that integrates modeling, optimization, and sensorless control of an all-electric heavy-duty robotic manipulator (HDRM) driven by electromechanical linear actuators (EMLAs). An EMLA model is formulated to capture motor electromechanics and direction-dependent transmission efficiencies, while a mathematical model of the HDRM, incorporating both kinematics and dynamics, is established to generate joint-space motion profiles for prescribed TCP trajectories. A safety-ensured trajectory generator, tailored to this model, maps Cartesian goals to joint space while enforcing joint-limit and velocity margins. Based on the resulting force and velocity demands, a multi-objective Non-dominated Sorting Genetic Algorithm II (NSGA-II) is employed to select the optimal EMLA configuration. To accelerate this optimization, a deep neural network, trained with EMLA parameters, is embedded in the optimization process to predict steady-state actuator efficiency from trajectory profiles. For the chosen EMLA design, a physics-informed Kriging surrogate, anchored to the analytic model and refined with experimental data, learns residuals of EMLA outputs to support force and velocity sensorless control. The actuator model is further embedded in a hierarchical virtual decomposition control (VDC) framework that outputs voltage commands. Experimental validation on a one-degree-of-freedom EMLA testbed confirms accurate trajectory tracking and effective sensorless control under varying loads.
\end{abstract}


\section{Introduction}
The escalating urgency of the climate crisis, underscored by the Paris Agreement \cite{unfccc2015paris} and the European Union’s 2035 ban on internal-combustion vehicles \cite{eu2035ban}, is accelerating the decarbonization of off-highway machinery and mobile working machines \cite{sorknaes2022electrification,bellocchi2020electrification}. Within this category, heavy-duty robotic manipulators (HDRMs) have traditionally relied on hydraulic linear actuators (HLAs) for their high power density in heavy-payload tasks \cite{7852495}. However, the multiple energy conversion stages from electrical generation to hydraulic pressure to mechanical motion introduce significant inefficiencies, with actuator efficiencies often below 50\% due to leakage, throttling losses, and auxiliary cooling demands \cite{manring2013efficiency,chipka2015efficiency}. Performance further degrades with fluid compressibility and viscosity variations, which impair response speed and positional accuracy under changing loads and temperatures \cite{pustavrh2023comparison}. In addition, frequent maintenance such as leak inspections, filter replacement, and oil top-ups increases downtime and costs, while spilled hydraulic fluid creates environmental and cleanup risks.  

Amid these challenges, and as cities mandate carbon-free public works by 2025 \cite{stokke2023procurement}, attention is shifting toward electromechanical linear actuators (EMLAs) \cite{bahari2025system}. Powered by permanent magnet synchronous motors (PMSMs) and precision screw transmissions, EMLAs eliminate fluid-based losses, enable regenerative braking during load-lowering cycles \cite{qu2023electrified}, and support integrated, high-bandwidth sensing essential for stability-guaranteed control. Furthermore, PMSM-driven EMLAs benefit from low-cogging-torque designs for ultra-smooth motion \cite{tootoonchian2016cogging} and provide accurate feedback for high-precision, fault-tolerant position sensing \cite{8854990,khajueezadeh2022reliability}.

From a control perspective, a wide range of strategies have been investigated for HDRMs, including genetic neural networks~\cite{li2018compliance}, nonlinear model predictive control~\cite{mononen2019nonlinear}, data-driven reinforcement learning~\cite{yao2023data}, adaptive neural network control~\cite{liang2024adaptive}, backstepping-based schemes~\cite{truong2023backstepping}, and virtual decomposition control (VDC)~\cite{hejrati2025orchestrated}. In parallel, numerous controllers have been proposed for PMSMs~\cite{dat2023advanced,niu2020robust}. However, the strong coupling between low-level actuator dynamics and high-level manipulator behavior underscores the need to analyze and design the control system as an integrated whole~\cite{barjini2025surrogate}. Within this context, modular frameworks such as VDC are particularly well-suited for evaluating and coordinating the performance of EMLAs in HDRMs~\cite{barjini2025surrogate}. Moreover, conventional PMSM torque and speed control requires multiple measurements that may be unavailable or impractical in many experimental setups~\cite{singh2017various}. Reducing dependence on such measurements through sensorless strategies therefore improves practicality by eliminating costly or failure-prone sensors.

Electrifying HDRMs with EMLAs and enabling robust, sensorless control requires a cohesive, end-to-end framework: one that captures EMLA dynamics and manipulator kinematics/dynamics, supports actuator configuration decisions for given trajectory demands, and ensures reliable control through sensorless strategies. To this end, the main contributions of this paper are summarized as follows:
\begin{enumerate}
  \item An EMLA model is obtained that integrates PMSM dynamics with direction-dependent screw-transmission efficiency, and is embedded through VDC into the HDRM joint-space kinematics and dynamics that are derived in this work.
  \item A safety-ensured trajectory generator is designed, tailored to the HDRM model, to map Cartesian trajectories into joint space and enforce joint limits and velocity margins to guarantee feasible motion execution under varying payloads.
  \item An EMLA configuration framework is formulated as a multi-objective NSGA-II problem over the motor, gearbox ratio, and screw lead, accelerated by a deep neural network model that delivers fast efficiency predictions of the actuator.
  \item A physics-informed Kriging (PIK) model is developed for the optimized EMLA to enable sensorless control with precise estimation of load-side force and velocity using only torque and angular velocity measurements, and integrated within the VDC framework. The unified controller is experimentally validated on a one-DOF EMLA testbed, demonstrating accurate trajectory tracking under varying loads.
\end{enumerate}

\section{Dynamic Modeling of the EMLA}
\label{sec:EMLA}
EMLAs provide precise motion conversion and high efficiency in all-electric HDRMs, which requires a unified model of PMSM electromechanics and screw-drive transmission losses. This section derives the PMSM voltage and torque equations, formulates the screw-drive kinematics with directional friction efficiency, and integrates both into a complete actuator dynamic model.
\subsection{PMSM Equations}
\label{subsec:PMSM}
Let \(\boldsymbol{i}_{abc}=[i_a,i_b,i_c]^\mathsf{T}\) and \(\boldsymbol{v}_{abc}=[v_a,v_b,v_c]^\mathsf{T}\) denote the phase currents and phase voltages, respectively. The stator resistance matrix is defined as \(\boldsymbol R_s=\mathrm{diag}(r_a,r_b,r_c)\), with inductance matrix \(\boldsymbol L_s\), and the flux‐linkage vector \(\boldsymbol\Psi_{abc}\). Here, \(P\) is the pole-pair count, \(\psi_f\) is the permanent‐magnet flux linkage per pole, and the flux linkage vector can be written as \(\boldsymbol\Psi_{abc}=\psi_f\,\frac{d}{dt}[\cos\,\theta_\gamma(t),\cos\,\theta_\delta(t),\cos\,\theta_\zeta(t)]^T\). In compact form, the stator phase voltage of PMSM is expressed as \eqref{eq:abc_voltage}.
\begin{equation}
\boldsymbol{v}_{abc} = \boldsymbol{R}_s \boldsymbol{i}_{abc} + \frac{d}{dt} \left(\boldsymbol{L}_s \boldsymbol{i}_{abc}\right)+\boldsymbol{\Psi}_{abc}.
\label{eq:abc_voltage}
\end{equation}

The Park transformation maps the stationary-frame vector \(\boldsymbol s_{abc}=[s_a,s_b,s_c]^\mathsf T\) into the rotating \(dq0\) frame as \(\boldsymbol s_{dq0}=\boldsymbol T_p\,\boldsymbol s_{abc}\), where \(\boldsymbol s_{dq0}=[s_d,s_q,s_0]^\mathsf T\), as detailed in \cite{abbas2024fast}. Accordingly, the voltage equations in the \(dq0\) frame are expressed as \eqref{eq:dq0_voltage}.

\begin{equation}
\begin{aligned}
\boldsymbol{v}_{dq0} &= \boldsymbol{T}_p \boldsymbol{R}_s \boldsymbol{T}^{-1}_p \boldsymbol{i}_{dq0} 
+ \boldsymbol{T}_p \frac{d}{dt} \left( \boldsymbol{L}_s \boldsymbol{T}^{-1}_p \right. \\
&\quad \left. \boldsymbol{i}_{dq0} \right) 
+ \boldsymbol{T}_p \boldsymbol{T}^{-1}_p \boldsymbol{\Psi}_{dq0}.
\end{aligned}
\label{eq:dq0_voltage}
\end{equation}

Assuming a balanced three-phase PMSM with equal stator resistances, the current dynamics and the electromagnetic torque are expressed as \eqref{eq:PMSM_dynamic} \cite{10816226}.
\begin{equation}
\left\{
\begin{aligned}
 \frac{di_d}{dt} &=\frac{1}{L_d} v_d-\frac{r_s}{L_d} i_d+\frac{L_q}{L_d} P \omega_m i_q , \\
 \frac{di_q}{dt} &=\frac{1}{L_q} v_q-\frac{r_s}{L_q} i_q-\frac{L_d}{L_q} P \omega_m i_d -\frac{P \psi_f \omega_m}{L_q}  ,\\
 \tau_e&=1.5 P \left[\psi_f i_q+\left(L_d-L_q\right) i_d i_q \right].
\end{aligned}
\right.
\label{eq:PMSM_dynamic}
\end{equation}
\begin{remark}
The PMSM phase voltages are generated by a voltage‐source inverter (VSI) using pulse‐width modulation (PWM). The actuator controller first computes the required \(dq\)-axis voltages \((v_{d_r},v_{q_r})\), which are converted to three‐phase duty cycles via the inverse Park transform. The PWM module then modulates these duty cycles to produce the switching signals \(S_1\)–\(S_6\), which yield the phase voltages \(\boldsymbol v_{abc}\) applied to the motor.
\end{remark}
\subsection{Transmission Mechanism and Dynamic Model}
An EMLA’s screw drive converts motor rotary motion into linear displacement via the ratio \(n=2\pi/\rho\), where \(\rho\) is the screw lead. Thread friction results in direction‐dependent efficiencies: \(\eta_f\) is forward efficiency when the motor overcomes friction and load, and \(\eta_b\) is backdriving efficiency when loads back-drive the screw. Defining the lead angle \(\phi=\arctan(\rho/(2\pi r_m))\) and friction angle \(\lambda=\arctan(\mu)\), the screw efficiency \(\eta_s\) is given by:
\begin{equation}
\eta_s^{\pm} =
\begin{cases}
  \displaystyle
  \eta_f=\frac{\tan\bigl(\phi(\rho)\bigr)}
       {\tan\bigl(\phi(\rho)+\lambda(\mu)\bigr)},
  & \dot{x}_L \ge 0,\\[12pt]
  \displaystyle
  \eta_b= \frac{\tan\bigl(\phi(\rho)-\lambda(\mu)\bigr)}
       {\tan\bigl(\phi(\rho)\bigr)},
  & \dot{x}_L < 0.
\end{cases}
\label{eq:efficiency_directional}
\end{equation}

Denoting gearbox efficiency by \(\eta_g\), the direction‐aware transmission efficiency model \(\eta_t^{\pm}\) in terms of gear ratio \(N_g\), \(\mu\), and \(\rho\) is expressed as \(\eta_t^{\pm} = \eta_g(N_g) \times \eta_s^{\pm} (\mu,\rho)\).
Accounting for Coulomb friction \(\tau_c\), viscous damping \(f_v\), and external load \(F_L\), the EMLA dynamics are given in \eqref{eq:EMLA_dynamic_p}–\eqref{eq:EMLA_dynamic_n}:
\begin{equation}
\left\{
\begin{aligned}
\ddot{x}_L = \frac{n N_g \big(\tau_e - \tau_c- \kappa_f (f_v \dot{x}_L + F_L)\big)}{M_t + n N_g \kappa_f J_m},\\  
\text{where} \quad \ddot{x}_L \geq 0 \quad \text{and} \quad \kappa_f = n N_g \eta_t^{+}.
\end{aligned}
\right.
\label{eq:EMLA_dynamic_p}
\end{equation}

\begin{equation}
\left\{
\begin{aligned}
\ddot{x}_L = \frac{\kappa_b \left(\tau_e - \tau_c -n N_g f_v \dot{x}_L\right) - F_L}{M_t +  n N_g \kappa_b J_m},\\
\text{where} \quad \ddot{x}_L  < 0 \quad \text{and} \quad \kappa_b = \frac{n N_g \eta_g} {\eta_b}. 
\end{aligned}
\right.
\label{eq:EMLA_dynamic_n}
\end{equation}
where $M_t$ and $J_m$ denote the total moving mass and the motor inertia, respectively.

\section{Mathematical Modeling of the Manipulator and Safety-Ensured Trajectory Generation}
\label{sec:manipulator_modeling}

Building on the EMLA model presented in Section~\ref{sec:EMLA}, we integrate manipulator and actuator dynamics via the VDC framework~\cite{zhu2010virtual} to derive the kinematic and dynamic equations, and then introduce a safety‐guaranteed trajectory planner that enforces each EMLA’s allowable operating range.

\subsection{Kinematic and Dynamic Equations of the Manipulator}
\label{subsec:manipulator_kinematics}

\begin{remark}
The detailed kinematic and dynamic equations are derived in Algorithms~\ref{alg:VDC_Kinematics} and~\ref{alg:VDC_Dynamics}. In the rest of this section, we only summarize the key preliminaries of the VDC approach.
\end{remark}

\begin{definition}
    Let $\{\mathbf{A}\}$ be a three-dimensional orthogonal coordinate system attached to a rigid body. The origin's linear velocity in frame $\{\mathbf{A}\}$ is denoted by ${}^{A}\boldsymbol{v} \in \mathbb{R}^3$, and its angular velocity by ${}^{A}\boldsymbol{\omega} \in \mathbb{R}^3$. These are combined into the six-dimensional linear/angular velocity vector ${}^{A}\boldsymbol{V} = \left[ {}^{A}\boldsymbol{v}^T, {}^{A}\boldsymbol{\omega}^T \right]^T \in \mathbb{R}^6$, commonly used in rigid body dynamics to represent translational and rotational motion in a unified form~\cite{zhu2010virtual}.
\end{definition}

Now consider another frame $\{\mathbf{B}\}$, also attached to the same rigid body. The linear/angular velocity vector transforms between frames $\{\mathbf{A}\}$ and $\{\mathbf{B}\}$ as:
\begin{equation}
\mathbf{{}^{B}}{\boldsymbol{V}} ={ }^{\mathbf{A}} \mathbf{U}_{\mathbf{B}}^T \mathbf{{}^{A}}{\boldsymbol{V}},
\label{eq:transformation_V}
\end{equation}
where ${}^{\mathbf{A}} \mathbf{U}_{\mathbf{B}} \in \mathbb{R}^{6 \times 6}$ is the spatial velocity transformation matrix defined as:
\begin{equation}
{ }^{\mathbf{A}} \mathbf{U}_{\mathbf{B}}=\left[\begin{array}{cc}
{ }^{\mathbf{A}} \mathbf{R}_{\mathbf{B}} & \mathbf{0}_{3 \times 3} \\
\left({ }^{\mathbf{A}} \mathbf{r}_{\mathbf{A B}} \times\right) { }^{\mathbf{A}} \mathbf{R}_{\mathbf{B}} & { }^{\mathbf{A}} \mathbf{R}_{\mathbf{B}}
\end{array}\right],
\label{Eq:Transformation_Matrix}
\end{equation}
with ${}^{\mathbf{A}} \mathbf{R}_{\mathbf{B}} \in \mathbb{R}^{3 \times 3}$ denoting the rotation matrix from frame $\{\mathbf{B}\}$ to $\{\mathbf{A}\}$, and ${}^{\mathbf{A}} \mathbf{r}_{\mathbf{A B}} = [r_{x}, r_{y}, r_{z}]^{T}$ the position vector from $\{\mathbf{A}\}$ to $\{\mathbf{B}\}$, expressed in $\{\mathbf{A}\}$. The associated skew-symmetric cross-product operator is:
\begin{equation}
\left({ }^{\mathbf{A}} \mathbf{r}_{\mathbf{A B}} \times\right)=\left[\begin{array}{ccc}
0 & -r_{\mathrm{z}} & r_{\mathrm{y}} \\
r_{\mathrm{z}} & 0 & -r_{\mathrm{x}} \\
-r_{\mathrm{y}} & r_{\mathrm{x}} & 0
\end{array}\right].
\label{Eq:X_Operator}
\end{equation}
In the VDC framework, the vector $\boldsymbol{V} \in \mathbb{R}^6$ is sequentially transformed from the base to the end-effector, ensuring consistent motion representation throughout the kinematic chain.

\begin{definition}
    Let $\mathbf{{}^{A}\boldsymbol{f}} \in \mathbb{R}^3$ and $\mathbf{{}^{A}\boldsymbol{\tau}} \in \mathbb{R}^3$ denote the force and moment vectors applied to a rigid body, both expressed in frame $\{\mathbf{A}\}$. The combined force/moment vector is given by $\mathbf{{}^{A}}{\boldsymbol{F}} = \left[ \mathbf{{}^{A}\boldsymbol{f}}^{T}, \mathbf{{}^{A}\boldsymbol{\tau}}^{T} \right]^{T} \in \mathbb{R}^6$. In the VDC framework, this vector is propagated from the end-effector to the base~\cite{zhu2010virtual}.
\end{definition}

Now, consider another frame $\{\mathbf{B}\}$ attached to the same body. The force/moment vector transforms between $\{\mathbf{A}\}$ and $\{\mathbf{B}\}$ as:
\begin{equation}
\mathbf{{}^{A}}{\boldsymbol{F}} ={ }^{\mathbf{A}} \mathbf{U}_{\mathbf{B}}{ }^{\mathbf{B}} {\boldsymbol{F}}.
\label{eq:transformation_force}
\end{equation}

The net force/moment vector acting on the body, expressed in $\{\mathbf{A}\}$ and denoted by ${}^{\mathbf{A}} \boldsymbol{F}^* \in \mathbb{R}^6$, is computed using:
\begin{equation}
\mathbf{M}_{\mathbf{A}} \frac{\textit{d}}{\textit{dt}}\! \left({ }^{\mathbf{A}} \boldsymbol{V}\right) + \mathbf{C}_{\mathbf{A}} \! ^{\mathbf{A}}\! \boldsymbol{V} + \mathbf{G}_{\mathbf{A}} = {}^{\mathbf{A}} \boldsymbol{F}^*,
\label{eq:net_force}
\end{equation}
where $\mathbf{M}_{\mathbf{A}} \in \mathbb{R}^{6 \times 6}$ is the spatial mass matrix, $\mathbf{C}_{\mathbf{A}} \in \mathbb{R}^{6 \times 6}$ represents Coriolis and centrifugal terms, and $\mathbf{G}_{\mathbf{A}} \in \mathbb{R}^{6}$ is the gravity vector, all in frame $\{\mathbf{A}\}$. From~\eqref{eq:net_force}, the dynamics can be expressed in regressor form:
\begin{equation}
    Y_{\mathbf{A}} \theta_{\mathbf{A}} = \mathbf{M}_{\mathbf{A}} \frac{\textit{d}}{\textit{dt}}\! \left({ }^{\mathbf{A}} \boldsymbol{V}\right) + \mathbf{C}_{\mathbf{A}} \! ^{\mathbf{A}}\! \boldsymbol{V} + \mathbf{G}_{\mathbf{A}},
\end{equation}
where $Y_{\mathbf{A}}({}^{\mathbf{A}}\dot{\mathcal{V}},\,{}^{\mathbf{A}}\mathcal{V}) \in \mathbb{R}^{6 \times 10}$ is the regressor matrix and $\boldsymbol{\theta}_{\mathbf{A}}(m,\,{}^{\mathbf{A}}\mathbf{r}_{\mathbf{AB}},\,\mathbf{I}_{\mathbf{A}})$ the corresponding inertial parameter vector~\cite{hejrati2022decentralized}.

\subsection{Safety-Ensured Trajectory Generation}
\label{subsec:trajectory_planning}

To operate the all-electric HDRM, a trajectory generator computes the joint velocities required to reach a user-defined end-effector position in Cartesian space, using filtered joint angles as input, as shown in Fig.~\ref{fig:manipulator_strucure}.

A fifth-order polynomial trajectory generator produces a smooth trajectory from the initial position $\mathbf{P}_{\text{start}}$ to the final position $\mathbf{P}_{\text{final}}$ over time $t_{\text{end}}$, yielding:
\begin{equation}
\left\{
\begin{aligned}
\mathbf{P_d}(t) &= [x_d, y_d, z_d]^T = \sum_{i=0}^{5} \mathbf{a_i} t^i, \\
\mathbf{\dot{P}_{d}}(t) &= [\dot{x}_d, \dot{y}_d, \dot{z}_d]^T = \sum_{i=1}^{5} i \mathbf{a_i} t^{i-1}.
\end{aligned}
\right.
\label{Eq:desired_trajectory_end}
\end{equation}
The coefficients $\mathbf{a_i} \in \mathbb{R}^3$ satisfy boundary conditions with zero initial/final velocities and accelerations. After $t > t_{\text{end}}$, the position is fixed at $\mathbf{P}_{\text{final}}$ and the velocity is set to zero.

Following the VDC method, the required end-effector velocity adds a correction term to the desired velocity:
\begin{equation}
    \mathbf{\dot{P}_r} = \mathbf{\dot{P}_d} + \lambda (\mathbf{P_d} - \mathbf{P}),
\end{equation}
where $\mathbf{\dot{P}_r} \in \mathbb{R}^3$ is the required velocity, $\mathbf{P_d}$ and $\mathbf{\dot{P}_d}$ are the desired position and velocity, $\mathbf{P}$ is the measured position, and $\lambda$ is a positive gain.

The Cartesian-to-joint velocity mapping is derived using the Jacobian based on the Denavit–Hartenberg (DH) convention:
\begin{equation}
    \dot{\boldsymbol{\Pi}} = \mathbf{J} \, \dot{\boldsymbol{\Theta}},
    \label{Eq:InverseKinematics}
\end{equation}
where $\dot{\boldsymbol{\Pi}} \in \mathbb{R}^6$ is the end-effector velocity vector, $\mathbf{J} \in \mathbb{R}^{6 \times 6}$ is the Jacobian, and $\boldsymbol{\Theta}$ is the joint angle vector. Given $\dot{\boldsymbol{\Pi}}_r = [\dot{x}_r, \dot{y}_r, \dot{z}_r, 0, 0, 0]^T$, the required joint velocities are:
\begin{equation}
    \dot{\boldsymbol{\Theta}}_r = \mathbf{J}^{-1} \, \dot{\boldsymbol{\Pi}}_r.
    \label{Eq:InverseKinematics2}
\end{equation}

To enforce joint-level safety, a soft-limiting algorithm scales each required joint velocity $\dot{\zeta}_{i_r}$ as it nears its limit, resulting in:
\begin{equation}
\left\{
\begin{aligned}
\dot{\zeta}_{i_r}^\text{safe} &= s_i \cdot \dot{\zeta}_{i_r}, \quad \text{for } i = 1, \dots, 6, \\
s_i &=
\begin{cases}
\frac{\zeta_i - \zeta_{\min,i}}{m_i}, & \text{if } \zeta_i \leq \zeta_{\min,i} + m_i \text{ and } \dot{\zeta}_{i_r} < 0, \\
\frac{\zeta_{\max,i} - \zeta_i}{m_i}, & \text{if } \zeta_i \geq \zeta_{\max,i} - m_i \text{ and } \dot{\zeta}_{i_r} > 0, \\
1, & \text{otherwise},
\end{cases}
\end{aligned}
\right.
\label{Eq:required_safe}
\end{equation}
where $s_i \in [0, 1]$, $\zeta_{\min,i}$ and $\zeta_{\max,i}$ are joint limits, and $m_i$ is the soft margin. This ensures smooth deceleration near limits, avoiding abrupt velocity changes.

\section{Optimization Framework}
\label{sec:optimization}
\subsection{Problem Definition and Surrogate Training}
The actuator-configuration optimization problem is formulated by selecting the PMSM, gearbox ratio, and screw lead that enable an EMLA to follow a prescribed HDRM joint-space force–velocity trajectory. The objectives are twofold: (i) maximize the steady-state efficiency of the EMLA, and (ii) minimize the nominal power rating of the PMSM. This yields a multi-objective problem solved using the Non-dominated Sorting Genetic Algorithm II (NSGA-II) \cite{deb2002fast}. NSGA-II is chosen for its ability to generate a diverse Pareto front without predefined weights and for its elitism mechanism, which preserves high-quality solutions across generations.  

Each motor \(i\) in the catalog is characterized by a nominal shaft power \(P_{N_i}\) and a rated efficiency \(\eta_{N_i}\). Let \(P_{\mathrm{req}}\) denote the mechanical power required at the load. A capable set of motors is defined by retaining only those indices \(i\) that satisfy
\(
\eta_{N_i} \,\eta_t^{\pm}(N_g,\mu,\rho)\, P_{N_i} \;\ge\; P_{\mathrm{req}}(F_{L_r}, \dot{x}_{L_r}),
\)
so that \(i \in \{1,\dots,n_c\}\) always corresponds to a PMSM capable of meeting the trajectory demands.  

To reduce computational cost during optimization, rather than repeatedly invoking the full Simulink EMLA model, a deep neural network (DNN) model is trained. The DNN maps EMLA parameters and load-side motion profiles to the predicted steady-state efficiency, enabling rapid evaluations within the optimization loop.
% To this end, we sample a grid of operating points and design variables \(F_L \in \{10^3,2\times10^3,\dots,3\times10^5\}\,\mathrm{N}\), \(\dot{x}_L\in \{0.025,\,0.05,\,0.075,\,0.10\}\,\mathrm{m/s}\), \(N_g \in \{1,\,5,\,10,\,15,\,20\}\), and \(\rho \in \{1,\,5,\,10,\,15,\,20,\,25\}\,\mathrm{mm}\).
% For each feasible \((\texttt{motor}_i,F_L,\dot{x}_L,N_g,\rho)\), we record efficiency of the EMLA from a Simulink run, using motor parameters \(\{r_s,L_d,L_q,\psi_f,P,J_m,I_{\max},U_N\}\) and transmission data \(\{N_g,\rho,D,\mu,J_L\}\) (\(\approx 46000\) samples). 
The resulting dataset is normalized and used to train a \([64,32,16,4]\)‐layer feedforward network, with a 70/15/15 train/validation/test split. The trained network achieves a mean absolute error below 2\%, reducing per‐query evaluation time from tens of seconds to under one millisecond. The offline training procedure is summarized in Algorithm~\ref{alg:surrogate_training}.
\begin{algorithm}[htb]
\caption{DNN Training for EMLA Efficiency}
\label{alg:surrogate_training}
\begin{algorithmic}[1]
  \Require PMSM parameters catalog \(\{\texttt{motor}_i\}_{i=1}^M\)
  \Require Grids: load force \(F_L\), linear velocities \(\dot{x}_L\), 
           gear ratios \(N_g\), screw leads \(\rho\), and \(i\)
  \Ensure Trained DNN surrogate \(\texttt{net}\) and normalizers \(\texttt{xSet}, \texttt{ySet}\)
  \State \(X \gets \varnothing\) and \(Y \gets \varnothing\)
  \For{\(i = 1\) to \(M\)}
      \ForAll{\(F\in F_L,\;v\in \dot{x}_L\)}
        \ForAll{\(G\in N_g,\;\ell\in \rho\)}
            \State \(P_{\rm avail} \gets   \eta_{N_i}\eta_t^{\pm}(N_g,\mu,\rho) \times P_{N_i}\)
            \If{\(P_{\rm avail} \leq F \cdot v\)} \State \textbf{continue} \EndIf
            \State build \(\texttt{params}\) from \(\texttt{motor}_i,\,G,\,\ell\)
            \State \(\eta \gets \Call{Simulink}{\texttt{params},F,v}\)
            \State append \([F,\,v,\texttt{params}]\) to \(X\)
            \State append \(\eta\) to \(Y\)
        \EndFor
      \EndFor
    \EndFor
  \State \([X_n,\texttt{xSet}]\gets \Call{mapminmax}{X}\) 
  \State \([Y_n,\texttt{ySet}]\gets \Call{mapminmax}{Y}\)
  \State \(\texttt{net} \gets \Call{fitnet}{[64,32,16,4],\,\texttt{`trainlm'}}\)
  \State \(\Call{train}{\texttt{net},\,X_n,\,Y_n}\)
    \State \Call{save}{\texttt{`EMLA\_DNN'.mat}, \texttt{net}, \texttt{xSet}, \texttt{ySet}}
\end{algorithmic}
\end{algorithm}

\begin{figure}[h]
  \centering
  \includegraphics[width=0.34\textwidth]{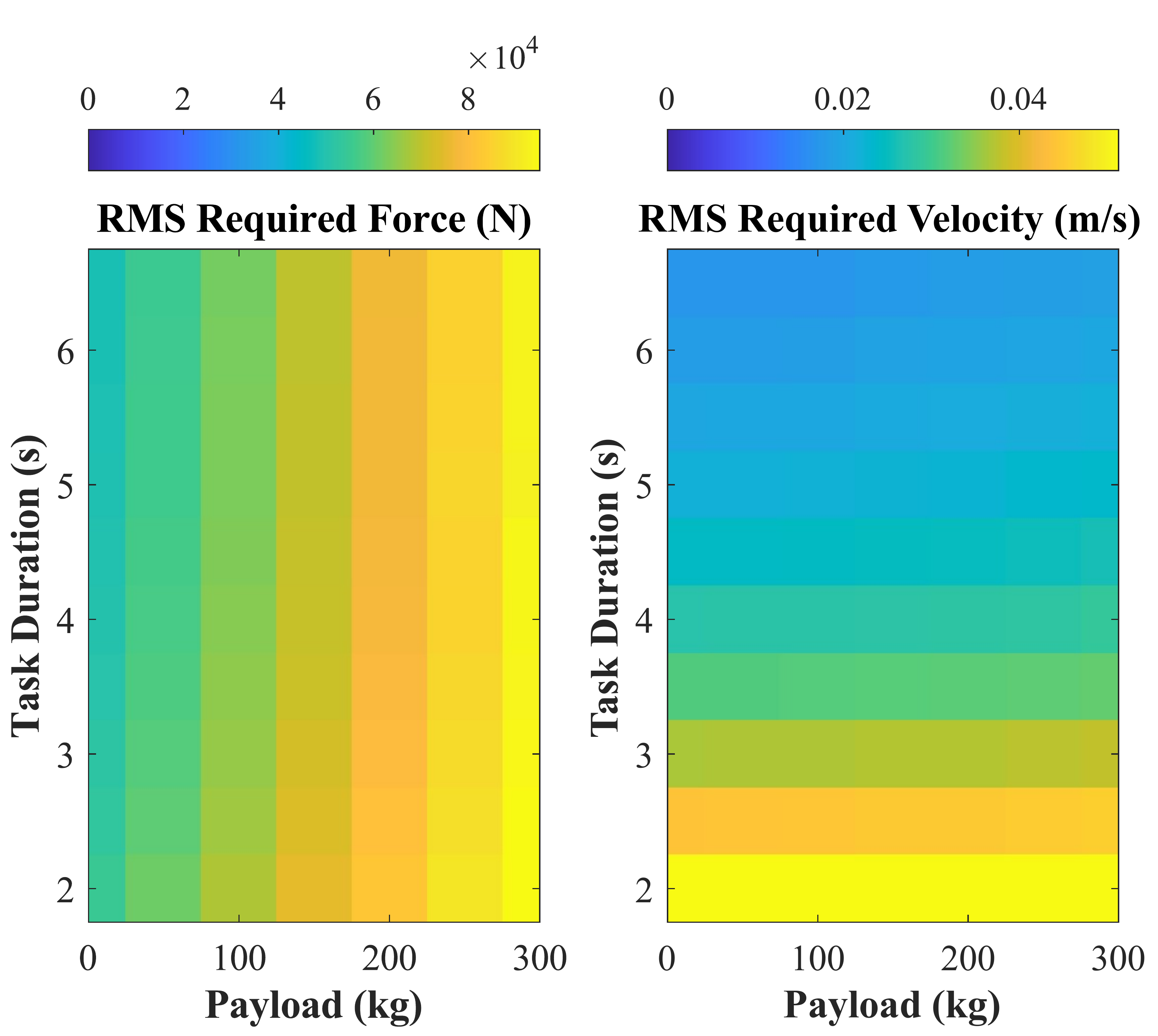}
  \caption{Required lift-joint force (N) and velocity (m/s) as functions of payload and task duration. These profiles, derived from the HDRM model, define the force–velocity demands that serve as inputs to the optimization framework.}
  \label{fig:req_force_velocity}
\end{figure}

Then, the manipulator’s Cartesian trajectories are mapped into joint space using the VDC‐derived HDRM kinematics and dynamics model (see Section~\ref{sec:manipulator_modeling}), yielding the required lift-joint force and linear velocity profiles, as shown in Fig.~\ref{fig:req_force_velocity}. These profiles \((F_{L_r},\dot x_{L_r})\), together with the PMSM catalog, serve as inputs to the optimization. Algorithm~\ref{alg:nsga2_emla_online} implements NSGA-II to solve the optimization problem in \eqref{eq:emla_optimization}, where the efficiency term \(\eta^{(i)}_{D\!N\!N}\) is predicted by the trained DNN model from Algorithm~\ref{alg:surrogate_training}, evaluated for the $i$th motor. The constraints ensure that each candidate actuator configuration satisfies the required power and motion demands. Finally, the decision vector is defined as \(\mathbf{x}=(\,i,\,N_g,\,\rho)\), where \(i\) is the motor index. The optimization problem is therefore formulated as \eqref{eq:emla_optimization}:
\begin{equation}
\label{eq:emla_optimization}
\begin{aligned}
  \min_{\,\mathbf{x}=(i,\,N_g,\,\rho)} 
    &\mathbf{f}(\mathbf{x})
    = \begin{pmatrix}
        -\,\eta^{(i)}_{D\!N\!N}\bigl(\texttt{params}, F_{L_r},\dot{x}_{L_r}\bigr)\\
        P_N^{(i)} (\tau_e,\,\omega_m)
      \end{pmatrix}
  \\[6pt]
  \text{s.t.}\quad
    &\eta^{(i)}_{D\!N\!N}\bigl(\texttt{params}, F_{L_r},\dot{x}_{L_r}\bigr)\;P_N^{(i)}(\tau_e,\,\omega_m)\\ &\ge P_{\mathrm{req}}(F_{L_r}, \dot{x}_{L_r}, \eta_s) 
      \\[3pt]
    &i \in \{1,\dots,n_{\rm c}\} \\[3pt]
    & N_{g, \min}\le N_g\le N_{g, \max}\\[3pt] 
    & \rho_{\min}\le \rho\le \rho_{\max}
\end{aligned}
\end{equation}

The optimization problem in \eqref{eq:emla_optimization} is solved using Algorithm~\ref{alg:nsga2_emla_online}, which outputs the Pareto front \(\{(\eta_j,P_j)\}\) and the best actuator configuration \((i^*,G^*,\ell^*)\).
\begin{algorithm}[htb]
\caption{NSGA-II Optimization for EMLA Configuration}
\label{alg:nsga2_emla_online}
\begin{algorithmic}[1]
  \Require Safety-ensured joint-space trajectory \(F_{L_r},\,\dot{x}_{L_r}\)
  \Require PMSM catalog \(\{\texttt{motor}_i\}_{i=1}^M\)
  \Require Trained DNN model \(\texttt{net}\) with normalizers \(\texttt{xSet}, \texttt{ySet}\) from Algorithm~\ref{alg:surrogate_training}
  \Ensure Pareto front \(\{(\eta_j,P_j)\}\) and best actuator configuration \((i^*,G^*,\ell^*)\)
  \State Define decision vector \(\mathbf{x} = (i,\,G,\,\ell)\)
  \State \(i\in\{1,\dots,M\}\), \(G\in[N_{g,\min},N_{g,\max}]\), \(\ell\in[\rho_{\min},\rho_{\max}]\)
  \State Compute required power $P_{\rm req} \gets {F_{L_r}\,\dot{x}_{L_r}}$
  \Function{ObjFun}{$\mathbf{x},\,F_{L_r},\,\dot{x}_{L_r}$}
    \State $\eta_{D\!N\!N} \gets \Call{\texttt{EffPred}}{\mathbf{x}, F_{L_r},\,\dot x_{L_r},\;\texttt{net},\texttt{xSet},\texttt{ySet}}$
    % \If{\(\eta_{D\!N\!N} \times \mathbf{x}[1].P_{N} < P_{\mathrm{req}}\)}
    %   \Return \([1e6]\) 
    % \EndIf
    \State \Return \(\left[-\eta_{DNN},\, P_N^{(i)}\right]\)
  \EndFunction
  \Function{NonlCon}{$\mathbf{x}$}
    \State $\tau_{\rm req}\gets F_{L_r} \ell \left(2\pi G \eta_t^{\pm}\right)^{-1}$
    \State $n_{\rm req}\gets 60\,G\,\dot{x}_{L_r} (\ell)^{-1}$
    \State $\eta_{D\!N\!N} \gets \Call{\texttt{EffPred}}{\mathbf{x}, F_{L_r},\,\dot x_{L_r},\;\texttt{net},\texttt{xSet},\texttt{ySet}}$
    \State $c_1\gets P_{\rm req} - \eta_{D\!N\!N}\;P_N^{(i)}$  
    \State $c_2\gets n_{\rm req} - n_n^{(i)}$
    \State $c_3\gets \tau_{\rm req} - \tau_n^{(i)}$
    \State \Return $\bigl[c_1,\,c_2,\,c_3\bigr]$
  \EndFunction
  \State Initialize population of decision vector \((\mathbf{x})\)
  \State \(\{\text{Obj},\,\text{Cons}\} \gets \)\texttt{gamultiobj}(\(\)ObjFun, NonlCon, \(\dots\))
  \For{each solution \(j\)}
    \State \(\eta_j \gets \text{-Obj}(j,1),\; P_j \gets \text{Obj}(j,2)\)
    \State Collect all non-dominated \(\{(\eta_j,P_j)\}\)
  \EndFor
  \State Identify best configuration \(\mathbf{x}_{j^*}=(i^*,G^*,\ell^*)\)
  \State \Return Pareto front \(\{(\eta_j,P_j)\}\) and best configuration \(\mathbf{x}_{j^*}\)
\end{algorithmic}
\end{algorithm}
\subsection{EMLA Selection and Results}
\begin{figure}[h]
  \centering
  \includegraphics[width=0.52\textwidth]{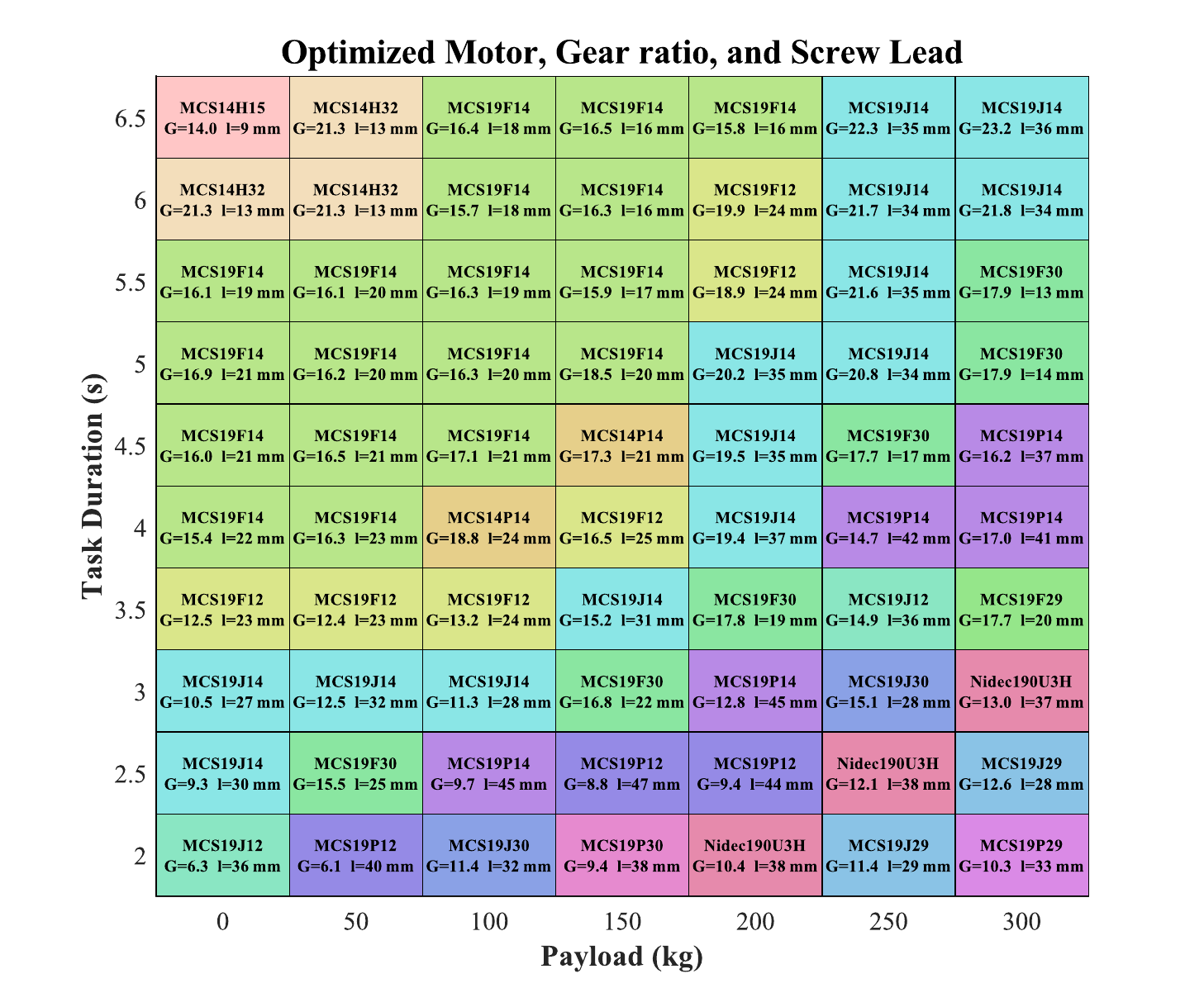}
  \caption{Optimized EMLA configuration grid where each cell indicates the selected motor, gear ratio, and screw lead for the corresponding payload–task duration pair.}
  \label{fig:Optimal_motor}
\end{figure}

The optimization is executed across the operational envelope of the HDRM, spanning a wide range of payloads and task durations. From the resulting Pareto front, each payload–task duration pair is mapped to its corresponding optimized actuator configuration, defined by the selected motor index \(i^*\), gearbox ratio \(G^*\), and screw lead \(\ell^*\), as shown in Fig.~\ref{fig:Optimal_motor}.

\begin{figure}[h]
  \centering
  \includegraphics[width=0.34\textwidth]{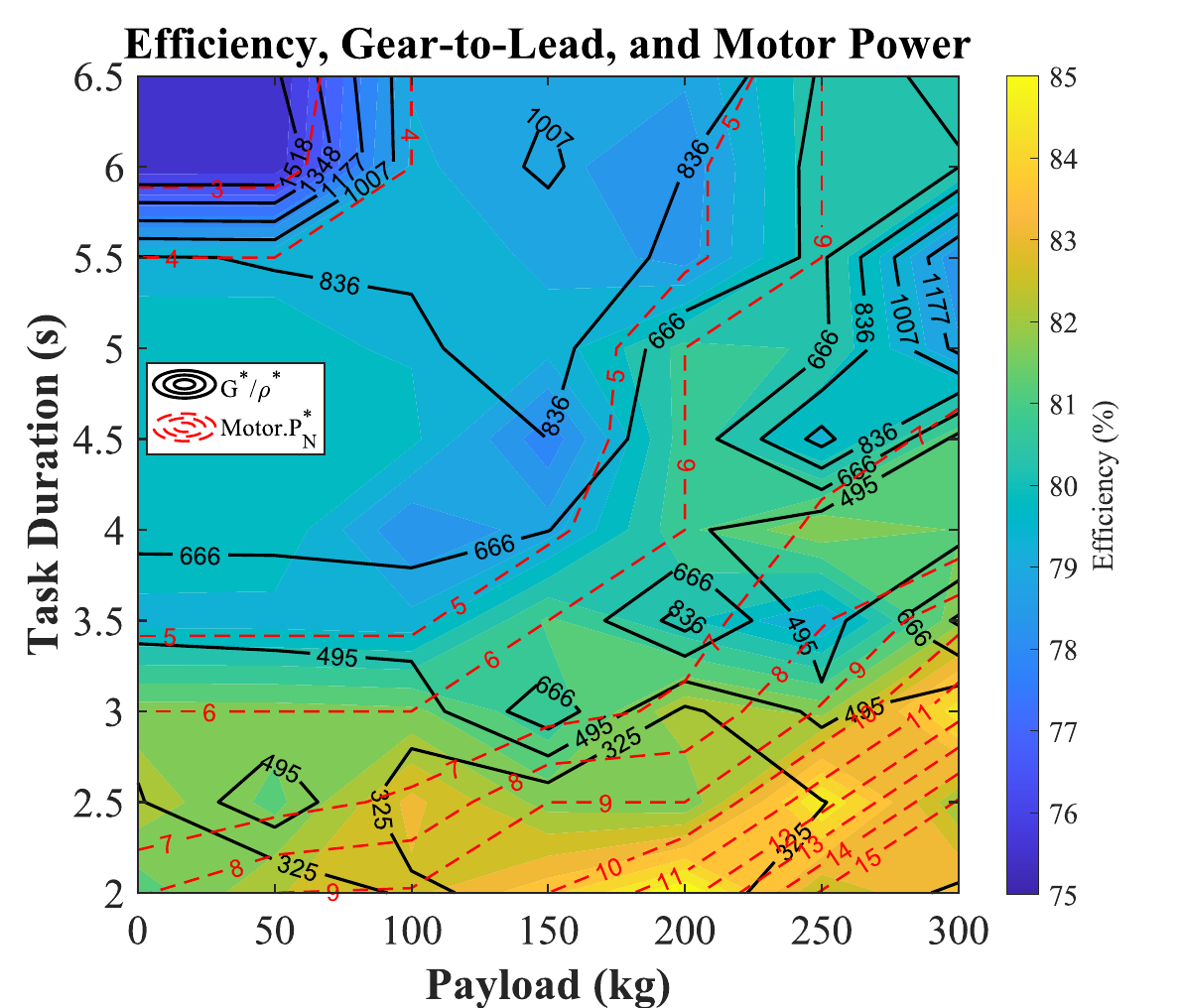}
  \caption{Efficiency contour of the optimized EMLA configuration (\%) with solid black isolines of gear-to-lead ratio and dashed red isolines of motor power (kW) as functions of payload and task duration.}
\label{fig:Efficiency_contour}
\end{figure}

The optimized designs exhibit clear trends in EMLA configuration across the operational envelope. For fast cycles, the NSGA-II selects the lowest gear‐to‐lead ratios, since excessive reduction or overly fine screw leads would unduly limit load‐side speed. As cycle times lengthen, the required actuator speed decreases, so the algorithm increases the gear-to-lead ratio, trading top-end speed for mechanical advantage. This reduces the motor’s power requirement and enables the selection of a smaller, lower-rated motor. Moreover, at any fixed cycle time, increasing payloads drive the optimal gear‐to‐lead ratio downward until the motor’s nominal power becomes insufficient; at that point NSGA-II elevates to a higher‐power motor rather than further changing the transmission parameters. Efficiency contours in Fig.~\ref{fig:Efficiency_contour} corroborate these design shifts, showing peak efficiency in the high‐payload, fast‐cycle regime and illustrating how larger motors at fast, heavy‐load conditions, operating close to their nominal point can outperform lighter‐load, slower‐cycle configurations with smaller motors. This is a noteworthy result for the heavy-duty electrification industry, as it demonstrates that upsizing the motor can enhance overall actuation efficiency under the most demanding operating scenarios.

Overall, by integrating NSGA-II with the trained DNN for EMLA efficiency prediction, the framework systematically exploits the torque–speed landscape by jointly balancing motor power and actuator efficiency. This yields actuator configurations that are optimally matched to each payload–cycle requirement. 
\section{Physics‐Informed Kriging Surrogate for the Optimized EMLA}
\label{sec:PIK}
This section develops a physics‐informed Kriging (PIK) surrogate for the optimized EMLA configuration. Given the HDRM’s specific payload–duration requirements, NSGA-II first identifies the best actuator configuration, which then serves as the basis for constructing the PIK model. The surrogate integrates analytic motor physics into a Gaussian‐process regression framework, enabling both high prediction accuracy and real‐time computational efficiency.
\begin{definition}
In the Gaussian process regression framework, let \(D\subseteq\mathbb{R}^d\) denote the input domain, with \(N\) observation locations \(
X = \{\,x^{(i)}\}_{i=1}^N\), where \(x^{(i)}\in D\),
and the associated scalar responses \(y = \bigl(y^{(1)},y^{(2)},\dots,y^{(N)}\bigr)^\top\), where \(y^{(i)}\in\mathbb{R}\).
The outputs are modeled as in \eqref{eq:vector_of_outputs}, with the Gaussian process prior given in \eqref{eq:gaussian_process}.
\begin{equation}
Y = \bigl(Y(x^{(1)}),\,Y(x^{(2)}),\,\dots,\,Y(x^{(N)})\bigr)^\top
\label{eq:vector_of_outputs}
\end{equation}
\begin{equation}
Y(\cdot)\sim\mathcal{GP}\bigl(\mu(\cdot),\,k(\cdot,\cdot)\bigr),
\label{eq:gaussian_process}
\end{equation}
where \(\mu(x)=\mathbb{E}\bigl[Y(x)\bigr]\) and
\(k(x,x')=\mathrm{Cov}\{Y(x),Y(x')\}\), while \(x,x'\in D\). The prior covariance matrix \(C\in\mathbb{R}^{N\times N}\) has entries \(C_{ij}=k\bigl(x^{(i)},x^{(j)}\bigr)\), and define the noise-aware training covariance \(K = C + \sigma_n^2 I\). The kernel hyperparameters are identified by maximizing the log-marginal likelihood
\begin{equation}
\left\{
\begin{aligned}
\ln L
&= -\tfrac12\,(y-\mu)^\top K^{-1}(y-\mu)
  - \tfrac12\ln\lvert K\rvert
  - \tfrac{N}{2}\ln(2\pi),\\
\mu
&= \bigl(\mu(x^{(1)}),\dots,\mu(x^{(N)})\bigr)^\top.
\end{aligned}
\right.
\end{equation}

For any test input \(x^*\in D\), the predictive distribution takes the form 
\begin{equation}
\left\{
\begin{aligned}
&Y(x^*)\mid X,y 
\sim \mathcal{N}\bigl(\hat y(x^*),\,\hat s^{2}(x^*)\bigr),\\
&\hat y(x^*) 
= \mu(x^*) + c(x^*)^\top K^{-1}(y-\mu),\\
&\hat s^{2}(x^*) 
= k(x^*,x^*) - c(x^*)^\top K^{-1}c(x^*),
\end{aligned}
\right.
\end{equation}
where the cross-covariance vector is defined as
\begin{equation}
c(x^*) = \bigl(k(x^{(1)},x^*),\,\dots,\,k(x^{(N)},x^*)\bigr)^\top.
\end{equation}
\end{definition}
\begin{remark}
  Here we take $d=2$, $D=\{(\tau_e,\omega_m)\}\subset\mathbb{R}^2$, and index inputs by
  $x^{(k)}=(\tau_e^k,\omega_m^k)$ for $k=1,\dots,N$.
\end{remark}

\subsection{Experimental Testbed and Data Collection}
\begin{figure}[h]
  \centering
  \includegraphics[width=0.45\textwidth]{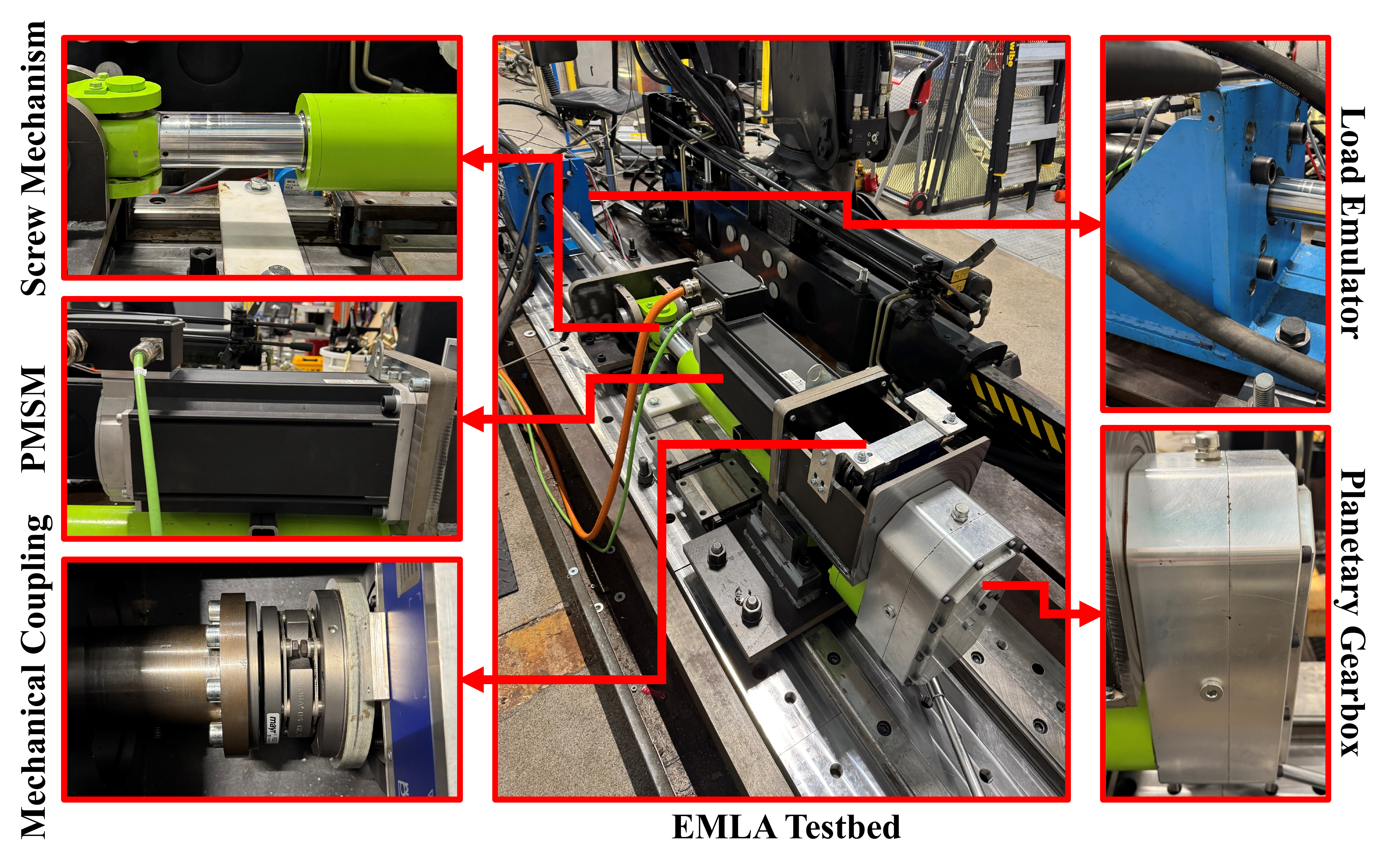}
  \caption{Experimental EMLA testbed.}
  \label{fig:experimental_testbed}
\end{figure}

The experimental setup employs an 11.6~\si{\kilo\watt}, 8-pole Nidec190U3H PMSM (380/480~V) coupled with a gearbox and screw drive to form the EMLA, as shown in Fig.~\ref{fig:experimental_testbed}. The actuator’s output is mechanically coupled to a secondary electrohydraulic cylinder that serves as a load emulator, generating dynamic, time-varying force profiles representative of HDRM lift-joint loading. Real-time signal acquisition and actuation feedback are handled via EtherCAT, ensuring precise synchronization and high-resolution monitoring.

Load forces ranged over \(F_L(t)\in[0,70]\)\,\si{\kilo\newton}, and linear velocities over \(\dot{x}_L(t)\in[0,0.07]\)\,\si{\meter\per\second}. All variables were sampled at \SI{1}{\kilo\hertz}. The raw time‐series $\{\tilde{F}_L(t),\,\tilde{\dot{x}}_L(t),\,\tilde{\eta}(t),\,\tilde{\tau}_e(t),\,\tilde{\omega}_m(t),\,\tilde{i}_{abc}(t)\}_{t=1}^{T}$ were post‐processed to identify characteristic operating points based on the desired velocity. The resulting measurements constitute the ground‐truth dataset for training and validating the PIK model. This selection strategy ensures the surrogate captures nonlinear effects such as saturation and friction across the full actuator envelope.
\begin{notation}
Let \((\tau_e,\omega_m)\) denote the PMSM's torque and angular velocity, and let [\(F_{L,\rm phys},\,\dot{x}_{L,\rm phys},\,\eta_{\rm phys},\,i_{abc,\rm phys}\)]
denote the corresponding analytic outputs.
The analytic model \texttt{M2L}, derived from Section~\ref{sec:EMLA}, maps the motor‐side outputs to EMLA's load‐side features:
\begin{equation}
(F_{L,\mathrm{phys}},\,\dot{x}_{L,\mathrm{phys}},\,\eta_{\mathrm{phys}},\,i_{abc,\mathrm{phys}})
= \texttt{M2L}(\tau_e,\omega_m).
\end{equation}
\end{notation}
This model serves as the physics-informed mean function around which the Gaussian-process residuals are learned.

\subsection{Physics-Informed GP Training and Evaluation}
\label{subsec:gp_training_results}
For each training sample \((\tau_e^k,\omega_m^k)\), we compute the analytic outputs via \texttt{M2L} and define the residuals as
\begin{equation}
\left\{
\begin{alignedat}{2}
  &\mathcal{R}_{F_L}^k      &{}={}& \tilde{F}_{L}^k   - F_{L,\mathrm{phys}}^k,\\
  &\mathcal{R}_{\dot{x}_L}^k&{}={}& \tilde{\dot{x}}_{L}^k - \dot{x}_{L,\mathrm{phys}}^k,\\
  &\mathcal{R}_{\eta}^k     &{}={}& \tilde{\eta}^k     - \eta_{\mathrm{phys}}^k,\\
  &\mathcal{R}_{i_{abc}}^k  &{}={}& \tilde{i}_{abc}^k  - i_{abc,\mathrm{phys}}^k.
\end{alignedat}
\right.
\end{equation}
We then assemble the raw feature matrix of motor-side variables,
\begin{equation}
  X_{\rm raw}
  \triangleq 
  \begin{bmatrix}
    \tau_e^1 & \omega_m^1 \\
    \tau_e^2 & \omega_m^2 \\
    \vdots   & \vdots     \\
    \tau_e^N & \omega_m^N
  \end{bmatrix}
  \in \mathbb{R}^{N\times 2},
\end{equation}
and compute column‐wise means and standard deviations
\begin{equation}
\left\{
\begin{aligned}
\mu_j &= \frac{1}{N}\sum_{k=1}^{N} X_{\mathrm{raw},\,j}^k,\\
\sigma_j &= \sqrt{\frac{1}{N-1}\sum_{k=1}^{N} \bigl(X_{\mathrm{raw},\,j}^k-\mu_j\bigr)^2},
\end{aligned}
\right.\quad j = 1,2,
\end{equation}
to normalize each input feature as
\begin{equation}
  X_{{\rm n},\,j}^k
  = \frac{X_{{\rm raw},\,j}^k - \mu_j}{\sigma_j},
  \quad k=1,\ldots,N,\; j=1,2.
\end{equation}
The vectors \(\mu,\sigma\in\mathbb{R}^2\) are stored to ensure identical preprocessing during prediction with the PIK model.
%
% \begin{definition}
% For PIK GP training, let \(\boldsymbol\mu=(\mu_1,\mu_2)\) and \(\boldsymbol\sigma=(\sigma_1,\sigma_2)\) be the column‐wise means and standard deviations used to normalize the raw inputs \(\tau_e\) and \(\omega_m\), respectively. Given \(N\) training samples \(\{(\tau_e^k,\omega_m^k),\,q_{\rm meas}^k\}\), for each \(k\) we form the normalized input as:
% %
% \begin{equation}
%   X^{k} = \bigl[\,(\tau_e^k-\mu_1)/\sigma_1,\;(\omega_m^k-\mu_2)/\sigma_2\bigr].
% \end{equation}
% %

% We then recover the raw torque and speed as:
% \begin{equation}
% \left\{
% \begin{aligned}
% \tau_{e,\mathrm{dn}}^k      &= X_{1}^k\,\sigma_1 + \mu_1,\\
% \omega_{m,\mathrm{dn}}^k    &= X_{2}^k\,\sigma_2 + \mu_2.
% \end{aligned}
% \right.
% \end{equation}
% \end{definition}
%

Four independent Gaussian‐process models, one for each output \(q\in\{F_L,\dot x_L,\eta,i_{abc}\}\), are trained with an ARD squared‐exponential kernel. The analytic map \(\texttt{M2L}\) serves as the mean function, while hyperparameters are tuned via quasi‐Newton optimization, as summarized in Algorithm~\ref{alg:PIK}.
\begin{algorithm}[]
\caption{PIK Surrogate Training for the EMLA}\label{alg:PIK}
\begin{algorithmic}[1]
  \Require $\{\tilde{F}_L(t),\,\tilde{\dot x}_L(t),\,\tilde{\eta}(t),\,\tilde{\tau}_e(t),\,\tilde{\omega}_m(t),\,\tilde{i}_{abc}(t)\}_{t=1}^T$
  \Ensure Trained model 
    $\displaystyle\texttt{gpPIKModel}$
    \For{$k=1,\dots,N$}
      \State Extract $(\tilde{F}^k_{L},\,\tilde{\dot{x}}^k_{L},\,\tilde{\tau}^k_{e},\,\tilde{\omega}^k_{m},\,\tilde{\eta}^k,\,\tilde{i}^k_{abc})$
      \State $(F_{L,\rm phys}^k,\,\dot{x}_{L,\rm phys}^k,\,\eta_{\rm phys}^k,\,i_{abc,\rm phys}^k)
        \gets \texttt{M2L}(\tau_e^k,\,\omega_m^k)$
    \EndFor
    \State $\mathcal R_q^k = \tilde{q}^k - q^{k}_{\rm phys}$
      for each $q\in\{F_L,\dot{x}_L,\eta,i_{abc}\}$
    \State $X_{\rm raw}[k,:]\gets [\,\tau_e^k,\;\omega_m^k\,]\in\mathbb{R}^{N\times2}$
    \State $\mu_j \gets \frac{1}{N} \sum_{k=1}^N X_{\rm raw}[k,j]$
    \State $\sigma_j \gets \sqrt{\frac{1}{N-1} \sum_{k=1}^N \bigl(X_{\rm raw}[k,j] - \mu_j\bigr)^2}$
    \State $X_{\rm norm}[k,j]\gets (X_{\rm raw}[k,j]-\mu_j)/\sigma_j$
    \State $m(X)\gets \texttt{M2L}\bigl(X[:,1]\cdot\sigma_1+\mu_1,\;
                         X[:,2]\cdot\sigma_2+\mu_2\bigr)$
    \For{$q\in\{F_L,\dot{x}_L,\eta,i_{abc}\}$}
      \State $\mathcal G_q 
             \gets \texttt{fitrgp}\bigl(X_{\rm norm},\,\mathcal R_q,\,m\bigr)$
    \EndFor
    \State $\texttt{gpPIKModel}\gets
        \{\mathcal G_{F_L},\,\mathcal G_{\dot x_L},\,\mathcal G_{\eta},\,\mathcal G_{i_{abc}},\,\mu,\,\sigma\}$
    \State \texttt{save(`gpPIKModel.mat',`gpPIKModel')}
  \State \Return \texttt{gpPIKModel}
\end{algorithmic}
\end{algorithm}

After fitting, the collection of GPs and normalization parameters is saved as \(
\texttt{gpPIK}
= \bigl\{\,
  \mathcal{G}_{F_L},\,
  \mathcal{G}_{\dot{x}_L},\,
  \mathcal{G}_{\eta},\,
  \mathcal{G}_{i_{abc}},\,
  \mu,\,
  \sigma
\,\bigr\}.
\)
Kernel length‐scales and noise variances are tuned via quasi-Newton maximization of the log-marginal likelihood (initialized using the data’s standard deviations). This PIK training scheme learns only the discrepancies between measured outputs and the inexpensive analytic map \(\texttt{M2L}\), yielding accurate, sub-millisecond predictions suitable for real-time control. 

Given \(P\) new motor–side pairs \(\{(\tau_e^p,\omega_m^p)\}_{p=1}^P\), the PIK surrogate computes load-side predictions and uncertainties as detailed in Algorithm~\ref{alg:predictPIK}.
\begin{algorithm}[]
\caption{PIK Surrogate Prediction}\label{alg:predictPIK}
\begin{algorithmic}[1]
  \Require Motor‐side inputs $\{(\tau_e^p,\omega_m^p)\}_{p=1}^P$, trained model 
  \(\texttt{gpPIK}=\{\mathcal G_q,\;\mu\in\mathbb{R}^2,\;\sigma\in\mathbb{R}^2\}\)
  \Ensure Output predictions $Y_{\rm pred}\in\mathbb{R}^{P\times4}$ 
  \For{$p=1,\dots,P$}
    \State $X_{\rm raw}[p,:]\gets [\,\tau_e^p,\;\omega_m^p\,]$
    \For{$j=1,2$}
      \State $X_{\rm n}[p,j]\gets \dfrac{X_{\rm raw}[p,j]-\mu_j}{\sigma_j}$
    \EndFor
  \EndFor
  \State $B_{\rm phys}\gets \texttt{M2L}(X_{\rm raw})$
  \For{$q=1$ \textbf{to} $4$}
    \State $(r_q,s_q)\gets \mathrm{predict}(\mathcal G_q,\,X_{\rm n})$
    \State $\Sigma(:,q)\gets s_q$
    \State $Y_{\rm pred}(:,q)\gets B_{\rm phys}(:,q)\;+\;r_q$
  \EndFor
  \State \Return $Y_{\rm pred}$
\end{algorithmic}
\end{algorithm}

We assess the surrogate’s predictive accuracy both qualitatively, through residual maps with uncertainty bands, and quantitatively, via RMSE on a hold-out test set. For each test sample \(p=1,\dots,P\), the surrogate outputs 
\(\{F_{L,\mathrm{pred}},\,\dot{x}_{L,\mathrm{pred}},\,\eta_{\mathrm{pred}},\,i_{abc,\mathrm{pred}}\}\) 
with predictive standard deviations 
\(\{\sigma_{F_L},\,\sigma_{\dot{x}_L},\,\sigma_{\eta},\,\sigma_{i_{abc}}\}\). 
Using measured values \(\tilde q^p\), we define the signed absolute error 
\(
  e_q^p = \tilde q^p - q^p_{\mathrm{pred}}
\)
and the signed percentage residual
\begin{equation}
  r_q^p = 100 \times \frac{\tilde q^p - q_{\mathrm{pred}}^p}{\tilde q^p},
  \quad
  q \in \{F_L,\dot{x}_L,\eta,i_{abc}\}.
  \label{eq:residual_equation}
\end{equation}

\begin{figure}[]
  \centering
  \includegraphics[width=0.4\textwidth]{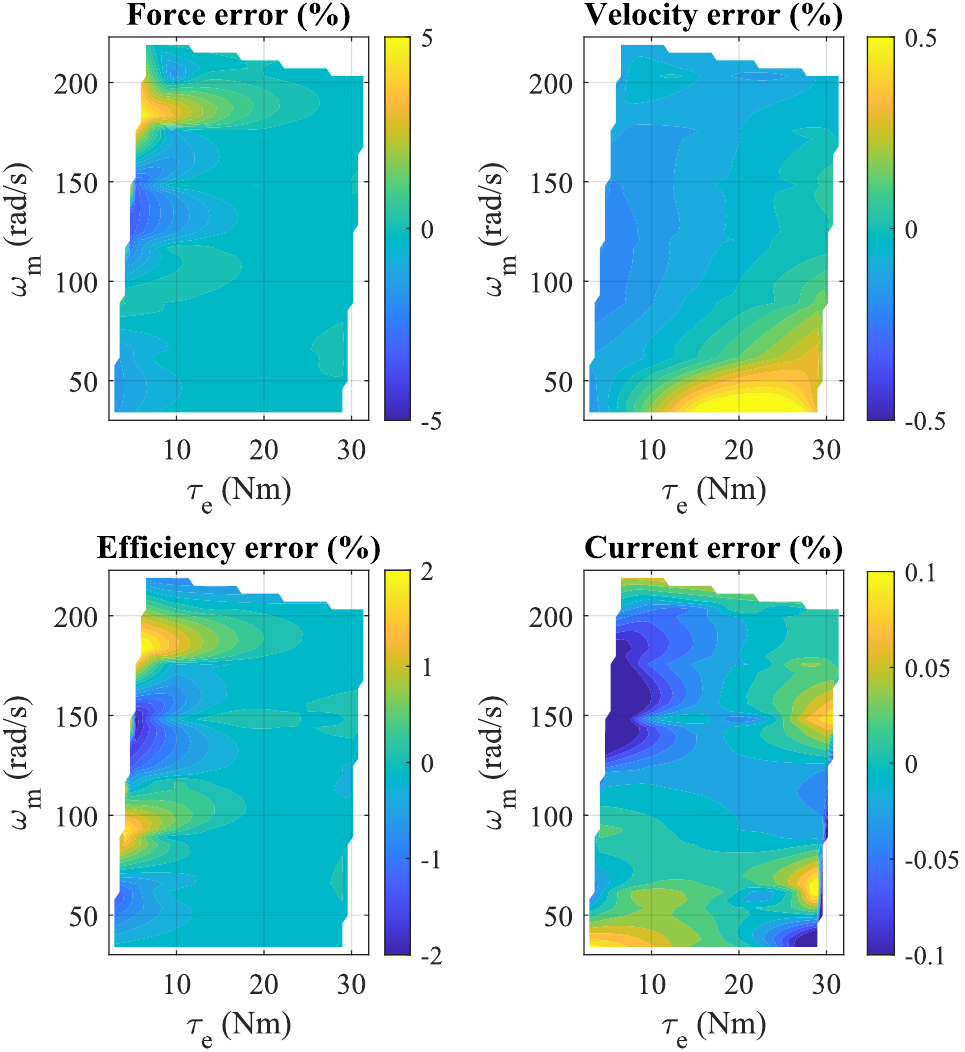}
  \caption{Residuals with predictive uncertainty (\(\pm2\sigma\)) across the torque–speed domain, confirming calibrated confidence intervals.}
  \label{fig:residuals}
\end{figure}

\begin{figure}[]
  \centering
  \includegraphics[width=0.4\textwidth]{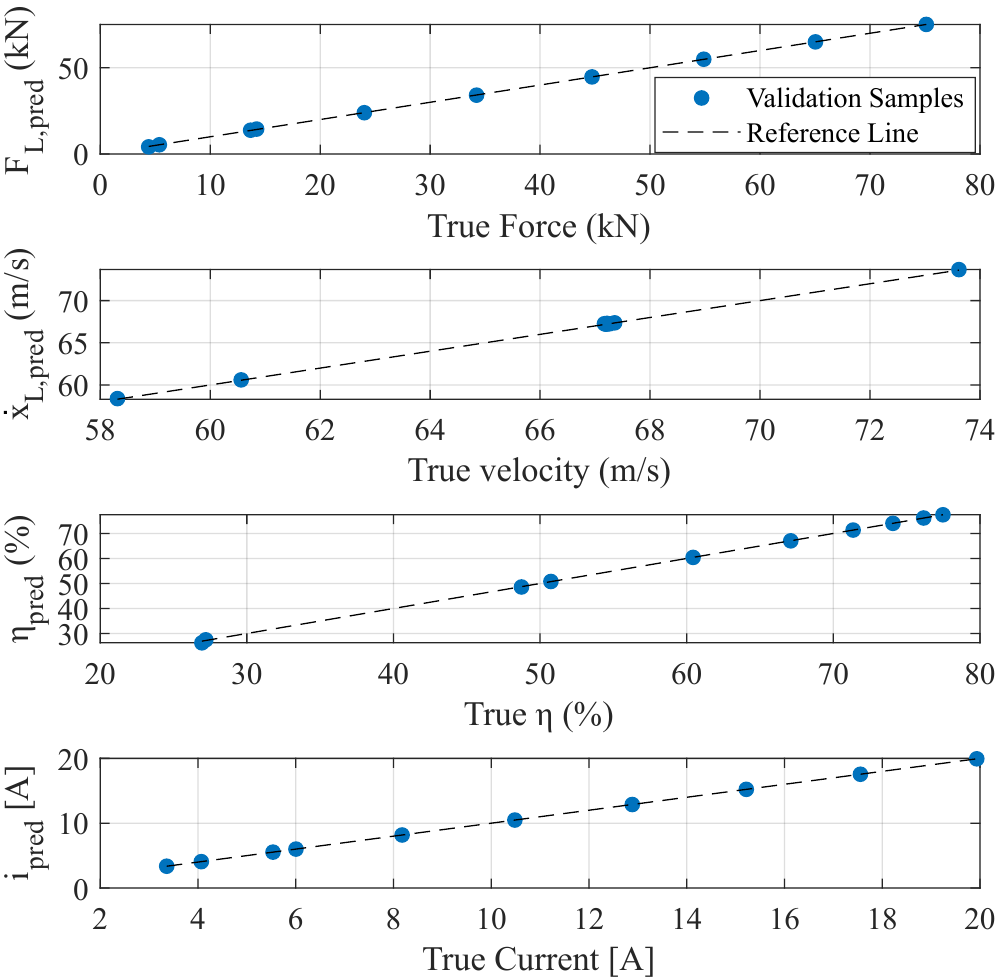}
  \caption{True versus predicted on the hold-out test set for force, velocity, efficiency, and current.}
  \label{fig:scatter}
\end{figure}

Residual maps with \(\pm2\sigma\) uncertainty bands are shown in Fig.~\ref{fig:residuals}, indicating that errors remain within the predicted bounds and that uncertainty increases smoothly in sparse data regions. On the hold-out test set, the surrogate achieves $\mathrm{RMSE}_{F_L}$= \SI{0.127}{\kilo\newton}, $\mathrm{RMSE}_{\dot x_L}$= \SI{0.001}{\meter\per\second}, $\mathrm{RMSE}_{\eta}\,$= \SI{0.23}{\percent}, and $\mathrm{RMSE}_{i_{abc}}$= \SI{0.0035}{\ampere}. Measured versus predicted values are plotted in Fig.~\ref{fig:scatter}, showing close alignment across all outputs.

\section{Controller Design}
\label{sec:manipulator_control}

In this section, the performance of the optimized EMLA configuration obtained in Section~\ref{sec:optimization} is evaluated to verify its capability to meet the required speed and force demands. Consequently, a controller must be designed for this system.

Building on the physics-informed Kriging (PIK) surrogate of the EMLA developed in Section~\ref{sec:PIK}, a hierarchical control framework is employed, as illustrated in Fig.~\ref{fig:manipulator_strucure}. At the high level, the VDC controller determines the required velocities and forces for each actuator, while at the low level, a dedicated controller generates the control voltages needed to achieve the commanded speed and force.

\subsection{High-Level Control}
\label{subsec:high-level_control}

\begin{figure*}[t]
\centering
\includegraphics[width=6in]{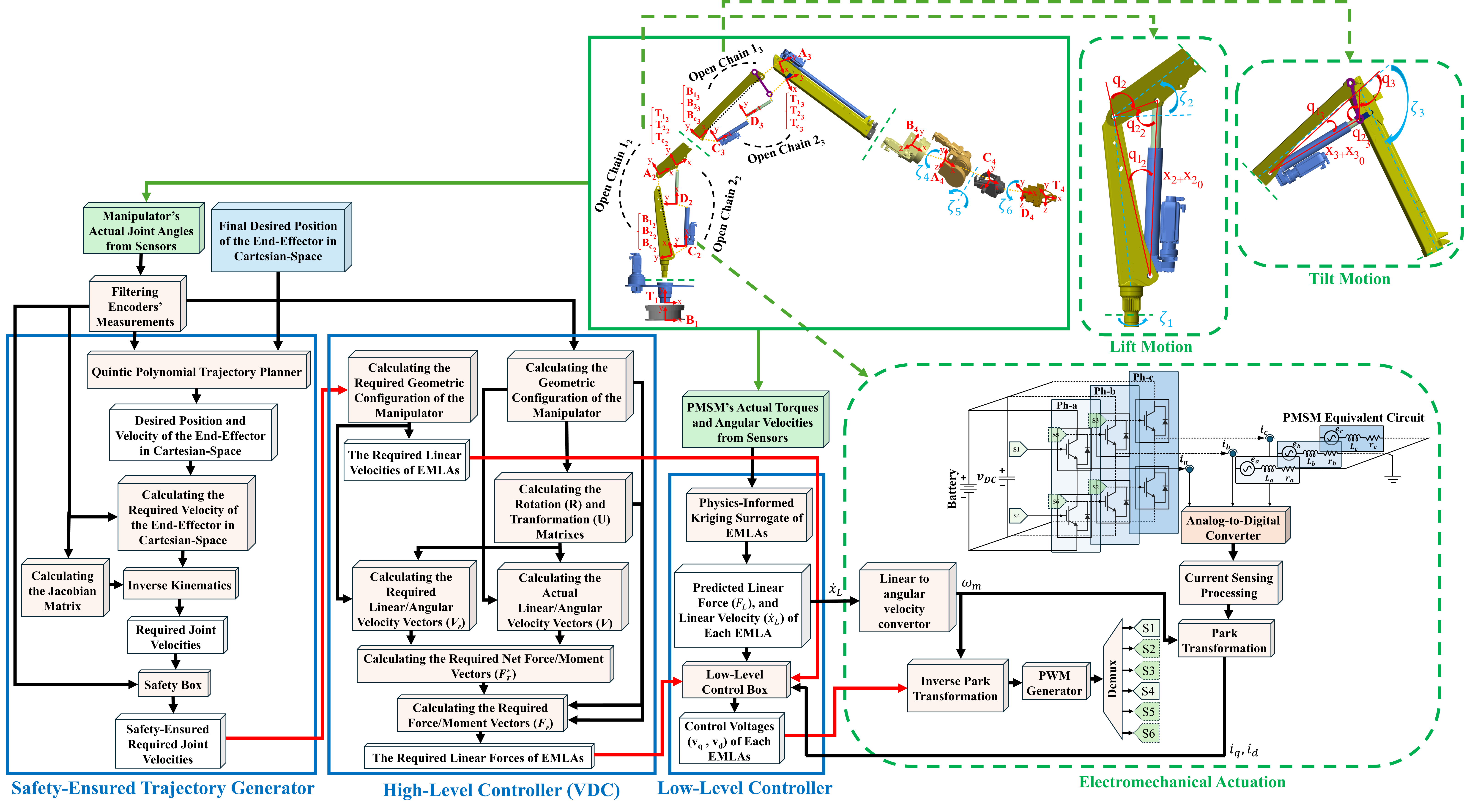}
\caption{Schematic of the Motion Planning and Control of the HDRM}
\label{fig:manipulator_strucure}
\end{figure*}

\begin{remark}
The high-level controller used here was introduced in~\cite{barjini2025surrogate} with full details; therefore, only a brief summary is provided.
\end{remark}

As shown in Fig.~\ref{fig:manipulator_strucure}, a modular adaptive VDC controller serves as the high-level controller. It computes the required linear velocities and forces for the EMLAs using safety-ensured joint velocity commands from the trajectory generator and filtered joint angle measurements.

Following the VDC framework~\cite{zhu2010virtual} and the procedure in~\cite{barjini2025surrogate}, the robot’s actual and required geometric configurations are determined from the filtered joint angles and the required joint velocities. Subsequently, the required linear velocities of all EMLAs ($\dot{x}_{L_r}$) are obtained, and the associated rotation and transformation matrices are calculated using~\eqref{Eq:Transformation_Matrix}.

Finally, the actual and required linear/angular velocity vectors expressed in any frame $\{\mathbf{A}\}$, denoted by $\mathbf{{}^{A}\boldsymbol{V}}, \mathbf{{}^{A}\boldsymbol{V}_r} \in \mathbb{R}^6$, are computed recursively from the base to the end-effector using Algorithm~\ref{alg:VDC_Kinematics}, as illustrated in Fig.~\ref{fig:manipulator_strucure}.

\begin{algorithm}[]
\caption{Systematic VDC-Based Kinematics}
\label{alg:VDC_Kinematics}
\begin{algorithmic}[1]
  \Require VDC model structure, joint variables ($\zeta_j$)
  \Ensure Linear/angular velocity vectors for all components

  \If{$j = 1$ \text{(base motion)}}
    \State Compute: $\left\{
    \begin{aligned}
    \mathbf{{}^{T_j}}{\boldsymbol{V}} &= \mathbf{y_{\tau}} \dot{\zeta}_j + { }^{\mathbf{B_j}} \mathbf{U}_{\mathbf{T_j}}^T \mathbf{{}^{B_j}}{\boldsymbol{V}} \\
    \mathbf{{}^{B_{c_{j+1}}}}{\boldsymbol{V}} &= { }^{\mathbf{T_j}} \mathbf{U}_{\mathbf{B_{c_{j+1}}}}^T \mathbf{{}^{T_j}}{\boldsymbol{V}}
    \end{aligned}
    \right.$

  \ElsIf{$j = 2$ \text{(lift motion)} or $j = 3$ \text{(tilt motion)}}
    \For{$j = 2$ to $3$}
      \State Chain $1_j$: $\left\{
      \begin{aligned}
      \mathbf{{}^{A_j}}{\boldsymbol{V}} &= \mathbf{z_{\tau}} \dot{q}_j + { }^{\mathbf{B_{1_j}}} \mathbf{U}_{\mathbf{A_j}}^T \mathbf{{}^{B_{1_j}}}{\boldsymbol{V}} \\
      \mathbf{{}^{T_{1_j}}}{\boldsymbol{V}} &= { }^{\mathbf{A_j}} \mathbf{U}_{\mathbf{T_{1_j}}}^T \mathbf{{}^{A_j}}{\boldsymbol{V}}
      \end{aligned}
      \right.$

      \State Chain $2_j$: $\left\{
      \begin{aligned}
      \mathbf{{}^{C_j}}{\boldsymbol{V}} &= \mathbf{z_{\tau}} \dot{q}_{1_j} + { }^{\mathbf{B_{2_j}}} \mathbf{U}_{\mathbf{C_j}}^T \mathbf{{}^{B_{2_j}}}{\boldsymbol{V}} \\
      \mathbf{{}^{D_j}}{\boldsymbol{V}} &= \mathbf{x_{f}} \dot{x}_j + { }^{\mathbf{C_j}} \mathbf{U}_{\mathbf{D_j}}^T \mathbf{{}^{C_j}}{\boldsymbol{V}} \\
      \mathbf{{}^{T_{2_j}}}{\boldsymbol{V}} &= \mathbf{z_{\tau}} \dot{q}_{2_j} + { }^{\mathbf{D_j}} \mathbf{U}_{\mathbf{T_{2_j}}}^T \mathbf{{}^{D_j}}{\boldsymbol{V}}
      \end{aligned}
      \right.$

      \State Motion constraints: $\left\{
      \begin{aligned}
      \mathbf{{}^{B_{1_j}}}{\boldsymbol{V}} &= \mathbf{{}^{B_{2_j}}}{\boldsymbol{V}} = \mathbf{{}^{B_{c_j}}}{\boldsymbol{V}} \\
      \mathbf{{}^{T_{1_j}}}{\boldsymbol{V}} &= \mathbf{{}^{T_{2_j}}}{\boldsymbol{V}} = \mathbf{{}^{T_{c_j}}}{\boldsymbol{V}}
      \end{aligned}
      \right.$

      \State Propagate: $\mathbf{{}^{B_{c_{j+1}}}}{\boldsymbol{V}}={} ^{\mathbf{T_{c_j}}}\mathbf{U}_{\mathbf{B_{c_{j+1}}}}^T\mathbf{{}^{T_{c_j}}}{\boldsymbol{V}}$
    \EndFor

  \ElsIf{$j = 4$ \text{(wrist motion)}}
    \State Compute: $\left\{
    \begin{aligned}
    \mathbf{{}^{A_j}}{\boldsymbol{V}} &= \mathbf{x_{\tau}} \dot{\zeta}_j + { }^{\mathbf{B_j}} \mathbf{U}_{\mathbf{A_j}}^T \mathbf{{}^{B_j}}{\boldsymbol{V}} \\
    \mathbf{{}^{C_j}}{\boldsymbol{V}} &= \mathbf{z_{\tau}} \dot{\zeta}_{j+1} + { }^{\mathbf{A_j}} \mathbf{U}_{\mathbf{C_j}}^T \mathbf{{}^{A_j}}{\boldsymbol{V}} \\
    \mathbf{{}^{D_j}}{\boldsymbol{V}} &= \mathbf{x_{\tau}} \dot{\zeta}_{j+2} + { }^{\mathbf{C_j}} \mathbf{U}_{\mathbf{D_j}}^T \mathbf{{}^{C_j}}{\boldsymbol{V}} \\
    \mathbf{{}^{T_j}}{\boldsymbol{V}} &= { }^{\mathbf{D_j}} \mathbf{U}_{\mathbf{T_j}}^T \mathbf{{}^{D_j}}{\boldsymbol{V}}
    \end{aligned}
    \right.$

  \EndIf

\end{algorithmic}
\end{algorithm}

Once the actual and required linear/angular velocity vectors are obtained, the required net force/moment vector in frame $\{\mathbf{A}\}$, denoted ${}^{\mathbf{A}} \boldsymbol{F}^*_r \in \mathbb{R}^6$, is computed as:
\begin{equation}
  {}^{\mathbf{A}} \boldsymbol{F}^*_r = Y_{\mathbf{A}} \hat{\theta}_{\mathbf{A}} + \mathbf{K_A} \left( {}^{\mathbf{A}} \boldsymbol{V}_r - {}^{\mathbf{A}} \boldsymbol{V} \right),
\label{Eq:Required_Force}
\end{equation}
where $\mathbf{K}_{\mathbf{A}} \in \mathbb{R}^{6 \times 6}$ is a positive-definite gain matrix, $Y_{\mathbf{A}}({}^{\mathbf{A}}\Dot{\mathcal{V}}_r,\,{}^{\mathbf{A}}\mathcal{V}_r) \in \mathbb{R}^{6 \times 10}$ is the regressor matrix, and $\hat{\boldsymbol{\theta}}_{\mathbf{A}}$ provides an estimate of the inertial parameter vector $\boldsymbol{\theta}_{\mathbf{A}}$ via the natural adaptation law~\cite{barjini2025surrogate}.

Following the VDC framework, the required force/moment vector in any frame $\{\mathbf{A}\}$, $\mathbf{{}^{A}\boldsymbol{F}_r} \in \mathbb{R}^6$, is then computed recursively from the end-effector to the base using Algorithm~\ref{alg:VDC_Dynamics}.

\begin{algorithm}[]
\caption{Systematic VDC-Based Dynamics}
\label{alg:VDC_Dynamics}
\begin{algorithmic}[1]
  \Require VDC model structure, required net force/moment vectors \((\boldsymbol{F}_r^*)\)
  \Ensure Required Force/moment vectors for all components

  \If{$j = 4$ \text{(wrist motion)}}
    \State Compute: $\left\{
    \begin{aligned}
    \mathbf{{}^{D_j}}{\boldsymbol{F}_r} &= \mathbf{{}^{D_j}}{\boldsymbol{F}_r^*} + { }^{\mathbf{D_j}} \mathbf{U}_{\mathbf{T_j}} \mathbf{{}^{T_j}}{\boldsymbol{F}_r}\\
    \mathbf{{}^{C_j}}{\boldsymbol{F}_r} &= \mathbf{{}^{C_j}}{\boldsymbol{F}_r^*} + { }^{\mathbf{C_j}} \mathbf{U}_{\mathbf{D_j}} \mathbf{{}^{D_j}}{\boldsymbol{F}_r}\\
    \mathbf{{}^{A_j}}{\boldsymbol{F}_r} &= \mathbf{{}^{A_j}}{\boldsymbol{F}_r^*} + { }^{\mathbf{A_j}} \mathbf{U}_{\mathbf{C_j}} \mathbf{{}^{C_j}}{\boldsymbol{F}_r}\\
    \mathbf{{}^{B_j}}{\boldsymbol{F}_r} &= \mathbf{{}^{B_j}}{\boldsymbol{F}_r^*} + { }^{\mathbf{B_j}} \mathbf{U}_{\mathbf{A_j}} \mathbf{{}^{A_j}}{\boldsymbol{F}_r}
    \end{aligned}
    \right.$

  \ElsIf{$j = 3$ \text{(tilt motion)} or $j = 2$ \text{(lift motion)}}
    \For{$j = 3$ to $2$}
      \State Compute:
      \[
      \begin{aligned}
      &\mathbf{{}^{B_{c_j}}}{\boldsymbol{F}_r} = \mathbf{{}^{B_{1_j}}}{\boldsymbol{F}_r^*} + { }^{\mathbf{B_{1_j}}} \mathbf{U}_{\mathbf{A_j}} \mathbf{{}^{A_j}}{\boldsymbol{F}_r^*} + { }^{\mathbf{B_{2_j}}} \mathbf{U}_{\mathbf{C_j}} \mathbf{{}^{C_j}}{\boldsymbol{F}_r^*} \\
      & + { }^{\mathbf{B_{2_j}}} \mathbf{U}_{\mathbf{C_j}} { }^{\mathbf{C_j}} \mathbf{U}_{\mathbf{D_j}} \mathbf{{}^{D_j}}{\boldsymbol{F}_r^*} + { }^{\mathbf{B_{1_j}}} \mathbf{U}_{\mathbf{A_j}} { }^{\mathbf{A_j}} \mathbf{U}_{\mathbf{B_{j+1}}} \mathbf{{}^{B_{j+1}}}{\boldsymbol{F}_r}
      \end{aligned}
      \]
    \EndFor

  \ElsIf{$j = 1$ \text{(base motion)}}
    \State Compute: $\mathbf{{}^{T_j}}{\boldsymbol{F}_r} = \mathbf{{}^{T_j}}{\boldsymbol{F}_r^*} + { }^{\mathbf{T_j}} \mathbf{U}_{\mathbf{B_{c_{j+1}}}} \mathbf{{}^{B_{c_{j+1}}}}{\boldsymbol{F}_r}.$
  \EndIf

\end{algorithmic}
\end{algorithm}

The required linear force for each EMLA ($F_{L_r}$) is subsequently determined from these results.

\subsection{Low-Level Control}
\label{subsec:low-level_control}

\begin{remark}
The low-level controller was originally introduced in~\cite{barjini2025surrogate}. To address practical constraints and reduce sensor costs, it was redesigned to operate without force and velocity sensors by leveraging the PIK surrogate from Section~\ref{sec:PIK}.
\end{remark}

The required velocity and force from Section~\ref{subsec:high-level_control} are tracked by each EMLA, as illustrated in Fig.~\ref{fig:manipulator_strucure}. First, based on the measured PMSM torque and angular velocity, the predicted load-side force \(F_{L,\mathrm{pred}} = Y_{\mathrm{pred}}(:,1)\) and predicted linear velocity \(\dot{x}_{L,\mathrm{pred}} = Y_{\mathrm{pred}}(:,2)\) are obtained using the surrogate PIK model (Algorithm~\ref{alg:predictPIK}). The required voltage inputs for the EMLA are then computed as~\cite{barjini2025surrogate}:
\begin{equation}
\left\{
\begin{aligned}
v_{d_r} &= r_s i_d + L_d \frac{d i_{d_r}}{dt} - L_q P \omega_m i_q + K_i (i_{d_r} - i_d), \\
v_{q_r} &= r_s i_q + L_q \frac{d i_{q_r}}{dt} + L_d P \omega_m i_d + P \psi_f \omega_m \\
&\quad + K_f (F_{L_r} - F_{L,\mathrm{pred}}) + K_v (\dot{x}_{L_r} - \dot{x}_{L,\mathrm{pred}}).
\end{aligned}
\right.
\label{eq:voltages_Control}
\end{equation}
Here, \(K_i\), \(K_f\), and \(K_v\) are positive scalar gains for current, force, and velocity tracking; \(i_d\) and \(i_q\) are the PMSM currents; and $F_{L_r}$ and $\dot{x}_{L_r}$ are the high-level controller’s required values. The feedforward terms account for electromechanical dynamics, while the feedback terms improve tracking accuracy and stability.
%The required current \(i_{d_r}\) is given by:
%\begin{equation}
%\frac{d}{dt} i_{d_r} = \frac{d}{dt} i_{d_d} + \lambda_i (i_{d_d} - i_d),
%\label{eq:current_d_required}
%\end{equation}
%where \(i_{d_d}\) is the desired direct-axis current and \(\lambda_i > 0\) is a control gain. Setting \(\frac{d}{dt} i_{d_d} = 0\) and \(i_{d_d} = 0\) recovers the conventional PMSM current control strategy~[].
The full hierarchical control, integrating the high-level VDC and low-level actuator controllers, is presented in Algorithm~\ref{alg:PIK-VDC}.
\begin{algorithm}[]
\caption{PIK-Enhanced VDC for All-Electric HDRM}\label{alg:PIK-VDC}
\begin{algorithmic}[1]
  \Require Required joint velocities \((\dot{\zeta}_{r_j})\), joint angle measurements \((\zeta_j)\), and PMSMs' torques and angular velocities \((\tau_{e_j}, \omega_{m_j})\)
  \Ensure Voltage commands for actuators \(\{v_{q_j}, v_{d_j}\}_{j=1}^6\)

  \Statex \textbf{Initialization:}
    \Statex \quad Load each joint’s PIK surrogate \(\texttt{gpPIKModel}_j\)
  \While{control active}
    \For{\(j=1\) \textbf{to} \(6\)}
    \State \textbf{High-Level Control:}
    \State Read joint states \(\zeta_j,\dot{\zeta}_j\)
    \State Compute \(\boldsymbol{V}\) using \(\dot{\zeta}_j\) (Algorithm~\ref{alg:VDC_Kinematics})
    \State Compute \(\boldsymbol{V}_r\) using \(\dot{\zeta}_{r_j}\) (Algorithm~\ref{alg:VDC_Kinematics})
    \State Compute the required linear velocities of EMLAs
    \State Compute
      \({}^{\mathbf{A}} \boldsymbol{F}^*_r = Y_{\mathbf{A}} \hat{\theta}_{\mathbf{A}} + \mathbf{K_A} \left( { }^{\mathbf{A}} \boldsymbol{V_r} - { }^{\mathbf{A}} \boldsymbol{V} \right)\)
    \State Compute \(\boldsymbol{F}_r\) (Algorithm~\ref{alg:VDC_Dynamics})
    \State Compute the required linear forces of EMLAs
      \State \textbf{Low-Level Control:}
      \State \(Y_{\mathrm{pred}}(j,:) = \texttt{PredPIK}(\tau_{e_j}, \omega_{m_j})\)
      \State \(\{F_{L,\mathrm{pred}_j},\dot{x}_{L,\mathrm{pred}_j}\} \gets Y_{\mathrm{pred}}(j,:)\)
      \State Compute \(v_{q_r}\) and \(v_{d_r}\) for each EMLA from \eqref{eq:voltages_Control}
      \State Send \(v_{q_r}\) and \(v_{d_r}\) via EtherCAT
    \EndFor
  \EndWhile
\end{algorithmic}
\end{algorithm}

\begin{remark}
The stability of the proposed modular control is analyzed in Section IV.C of~\cite{barjini2025surrogate} using Lyapunov theory, and its robustness to parametric uncertainties is discussed in Section V.A.
\end{remark}

\section{Experimental Results}
\label{sec:results}

\begin{figure}[h]
  \centering
  \includegraphics[width=0.48\textwidth]{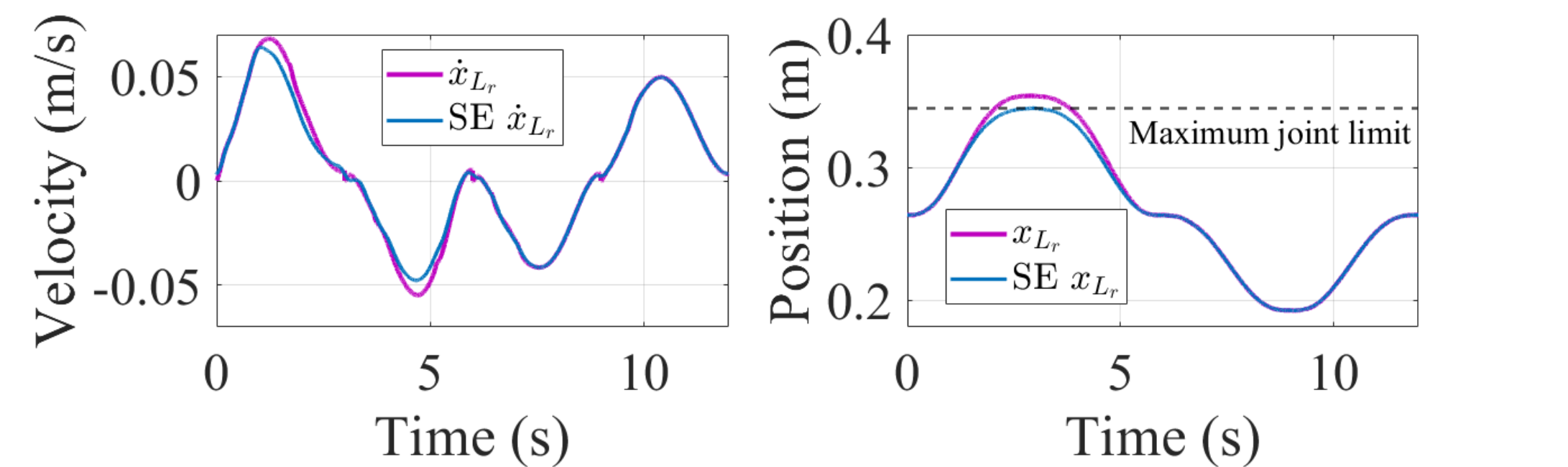}
  \caption{Required position and velocity, in the joint-space, before and after applying safety constraints.}
  \label{fig:Normal_SE}
\end{figure}

This section aims to: (i) assess the optimized EMLA configuration in meeting speed and force requirements, (ii) validate the safety-ensured (SE) trajectory generator, and (iii) compare the proposed sensorless (SL) VDC with the measurement-feedback (MF) VDC. To achieve these objectives, trajectories are designed to demand maximum velocity and reach joint limits, with payloads ranging from $0$ to $300\,\text{kg}$. Fig.~\ref{fig:Normal_SE} shows the required position and velocity before and after applying the safety constraints, as introduced in Section~\ref{subsec:trajectory_planning}.

\begin{figure}[]
  \centering
  \includegraphics[width=0.48\textwidth]{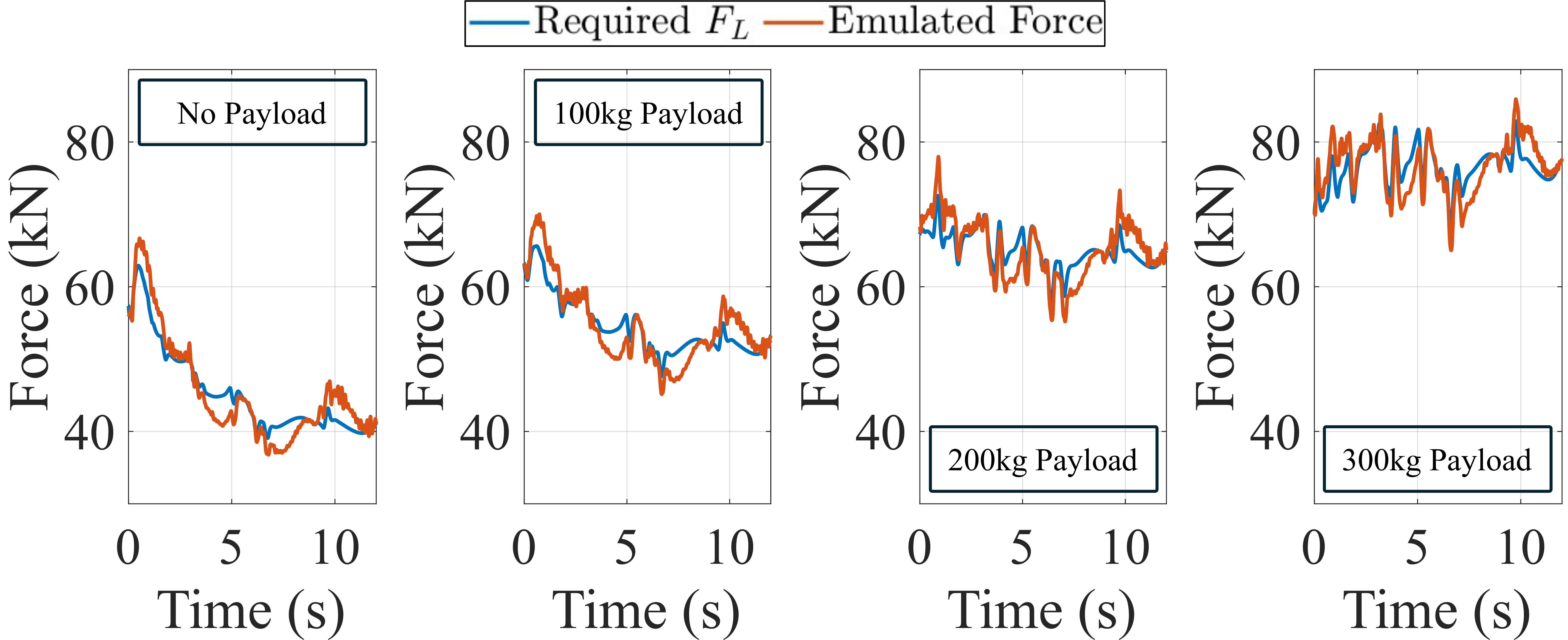}
  \caption{Comparison of the required force $F_{L_r}$ and the emulated force for different payloads.}
  \label{fig:forces}
\end{figure}

\begin{figure}[]
  \centering
  \includegraphics[width=0.48\textwidth]{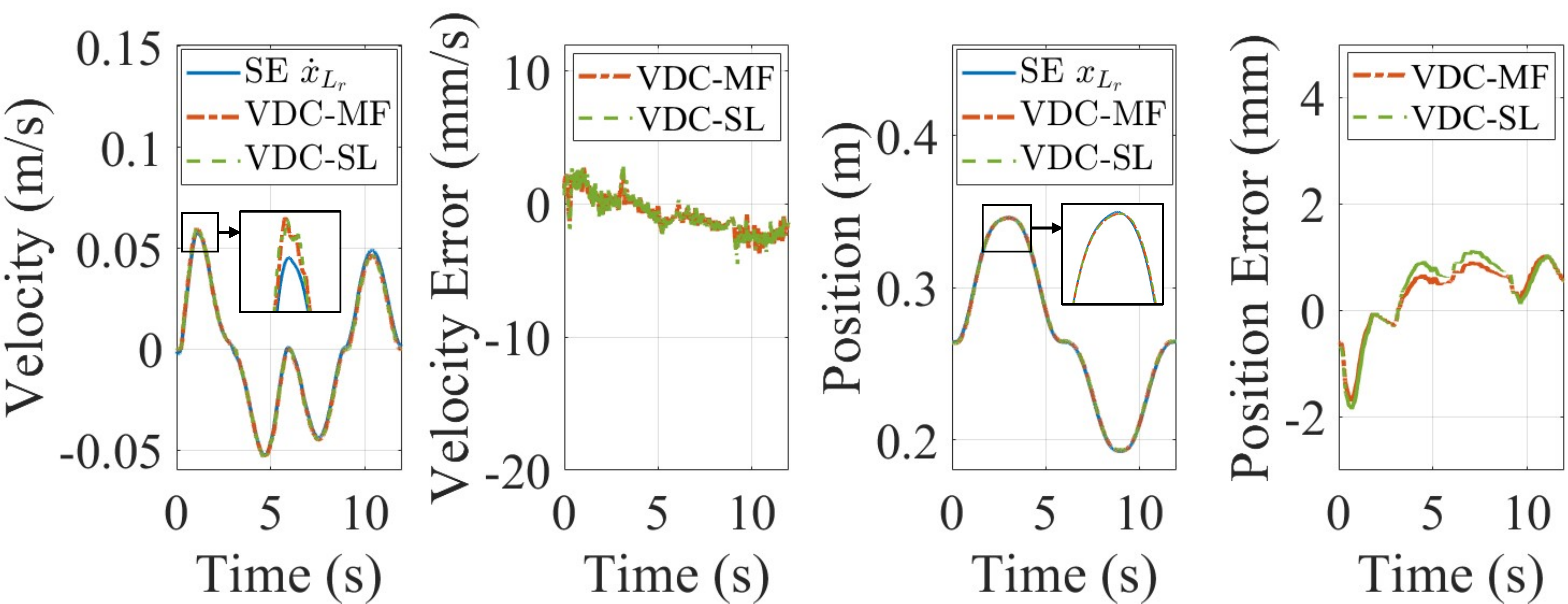}
  \caption{Tracking performance and errors under no payload condition.}
  \label{fig:0kg}
\end{figure}

\begin{figure}[]
  \centering
  \includegraphics[width=0.48\textwidth]{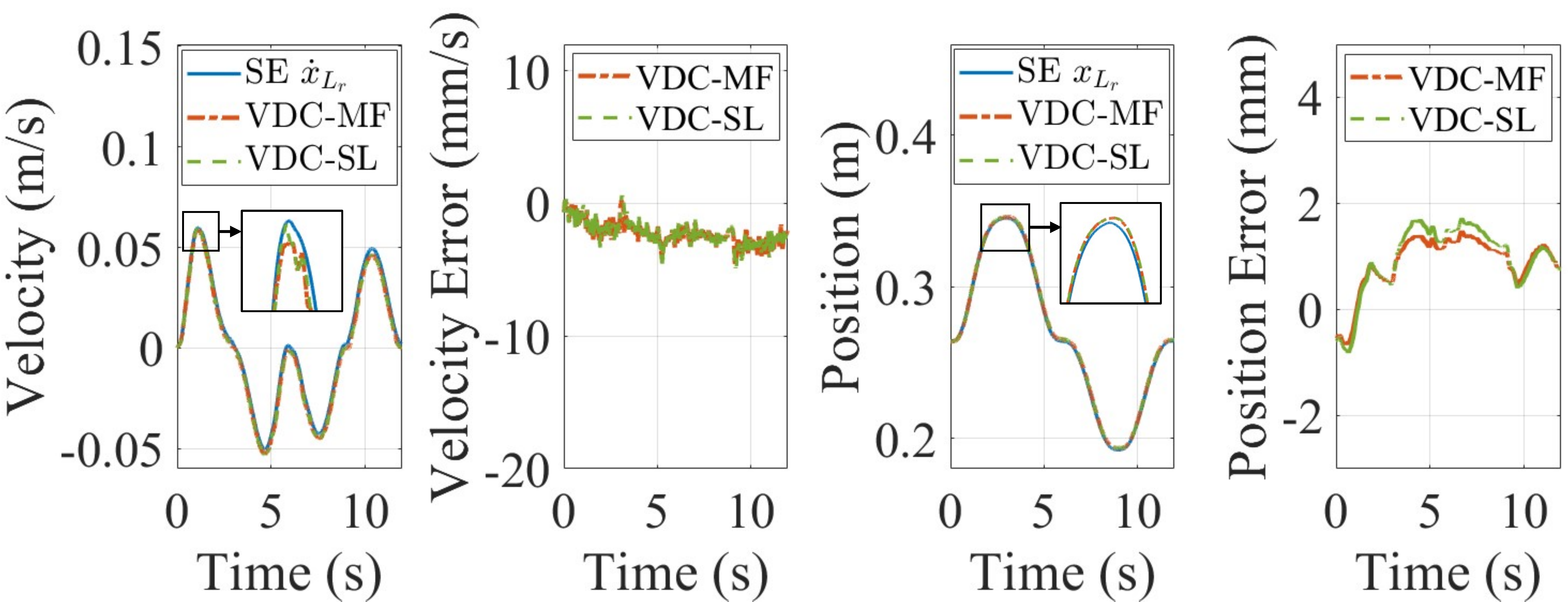}
  \caption{Tracking performance and errors under a 100 kg payload.}
  \label{fig:100kg}
\end{figure}

\begin{figure}[]
  \centering
  \includegraphics[width=0.48\textwidth]{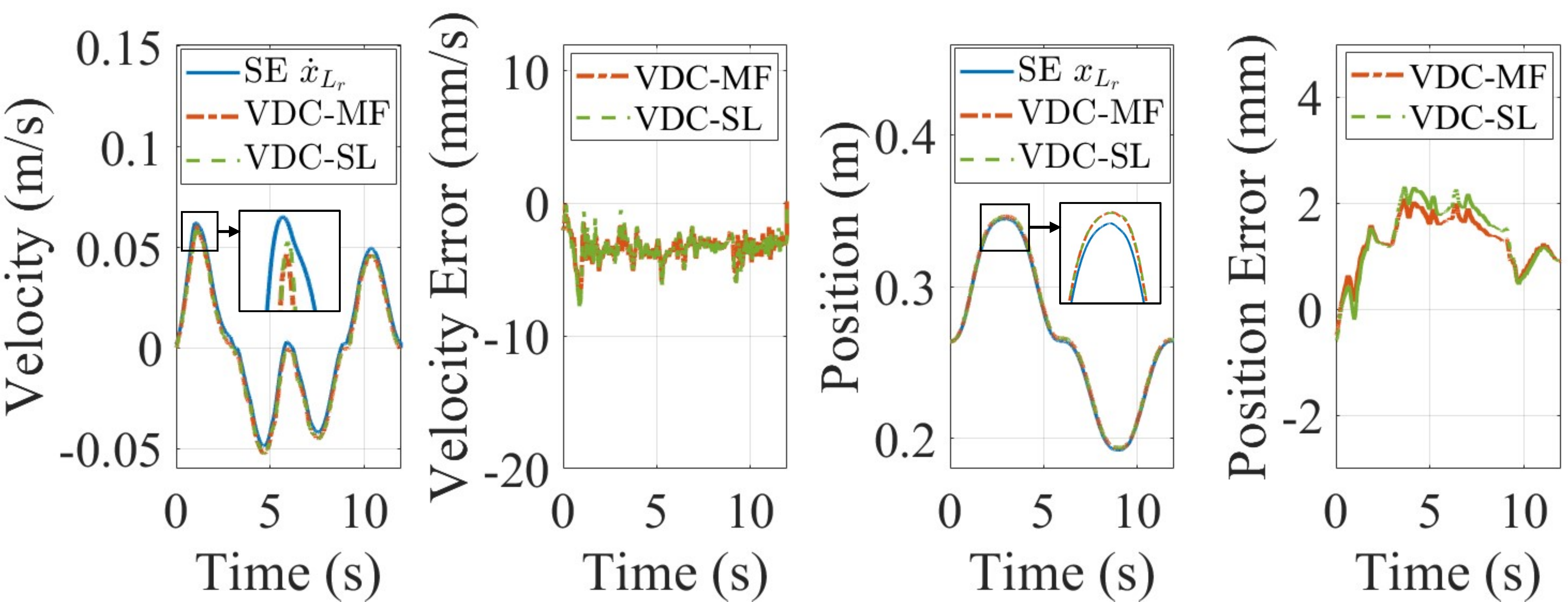}
  \caption{Tracking performance and errors under a 200 kg payload.}
  \label{fig:200kg}
\end{figure}

\begin{figure}[]
  \centering
  \includegraphics[width=0.48\textwidth]{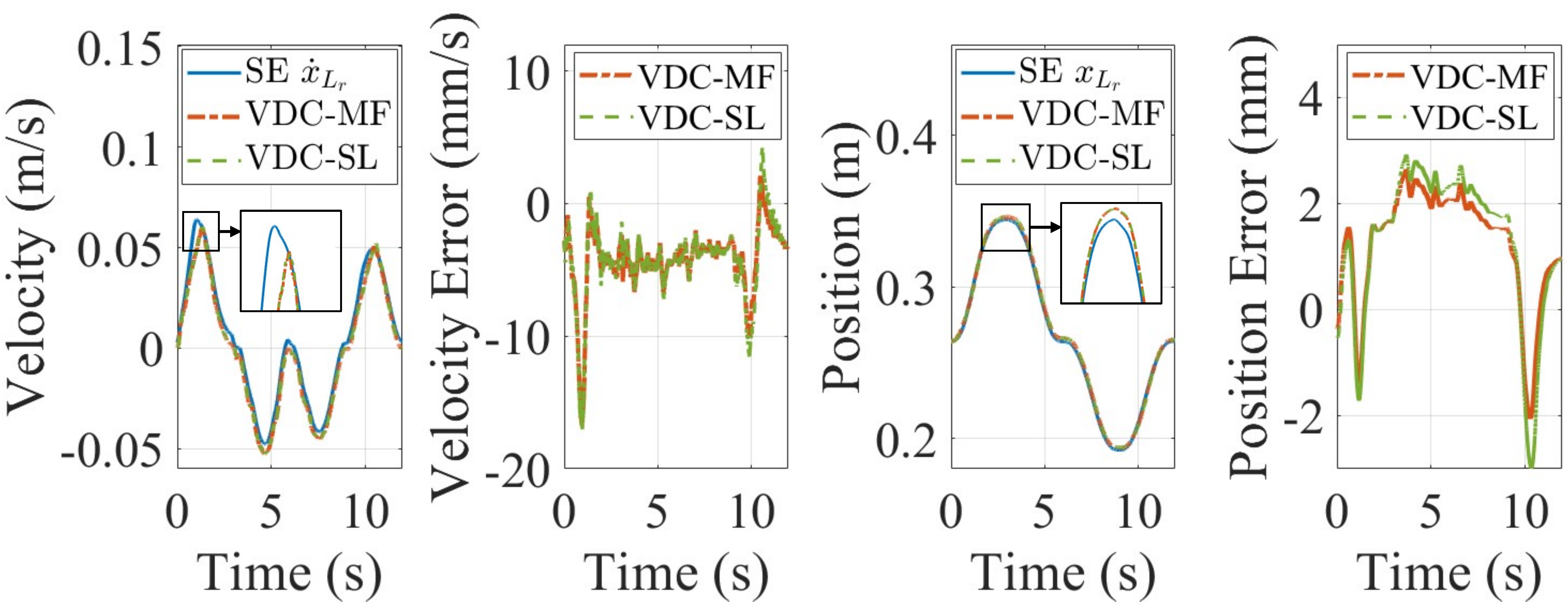}
  \caption{Tracking performance and errors under a 300 kg payload.}
  \label{fig:300kg}
\end{figure}

The results indicate that the SE trajectory effectively constrains joint motion within the safe range, ensuring that the maximum joint limit is not exceeded. 
Fig.~\ref{fig:forces} presents the required and emulated linear forces for SE trajectory tracking with payloads ranging from $0$ to $300\,\text{kg}$. Subsequently, using the SE trajectory generator, Figs.~\ref{fig:0kg}--\ref{fig:300kg} compare the position and velocity profiles of the VDC-SL and VDC-MF controllers.

It is observed that the VDC-SL accurately tracked position, velocity, and force, demonstrating robustness without linear velocity and force measurements. A quantitative comparison between the VDC-SL and VDC-MF is presented in Table~\ref{tab:errors}, using root mean square (RMS) and maximum (Max) tracking errors for position and velocity.
\begin{table}[h]
  \centering
  \begin{threeparttable}
  \caption{Tracking errors of VDC-MF and VDC-SL under different payload conditions}
    \label{tab:errors}
    \begin{tabular}{c|c|cc|cc}
      \hline
      \multirow{2}{*}{\textbf{Payload}} & \multirow{2}{*}{\textbf{Metric}} & \multicolumn{2}{c|}{\textbf{VDC-MF Errors}} & \multicolumn{2}{c}{\textbf{VDC-SL Errors}} \\
      \cline{3-6}
       &  & \textbf{Position} & \textbf{Velocity} & \textbf{Position} & \textbf{Velocity} \\
      \hline
      \multirow{2}{*}{0 kg}   & RMS & 0.72 & 1.56 & 0.82 & 1.63 \\
                              & Max & 1.71 & 3.71 & 1.86 & 4.50 \\
      \hline
      \multirow{2}{*}{100 kg} & RMS & 1.03 & 2.39 & 1.16 & 2.41 \\
                              & Max & 1.49 & 4.57 & 1.73 & 4.88 \\
      \hline
      \multirow{2}{*}{200 kg} & RMS & 1.39 & 3.45 & 1.52 & 3.48 \\
                              & Max & 2.10 & 7.19 & 2.32 & 7.86 \\
      \hline
      \multirow{2}{*}{300 kg} & RMS & 1.66 & 4.97 & 1.89 & 5.14 \\
                              & Max & 2.65 & 16.74 & 3.01 & 17.03 \\
      \hline
    \end{tabular}
    \begin{tablenotes}
      \footnotesize
      \vspace{1mm}
      \item[- Position errors are expressed in millimeters (mm).] 
      \item[- Velocity errors are expressed in millimeters per second (mm/s).] 
    \end{tablenotes}
  \end{threeparttable}
\end{table}
\normalsize

Table~\ref{tab:errors} indicates that both controllers exhibit increasing errors with larger payloads, highlighting the influence of HDRM inertia. The sensorless controller (VDC-SL) shows slightly higher RMS and maximum errors compared to the measurement-feedback controller (VDC-MF); however, all errors remain within acceptable limits, confirming accurate position, velocity, and force tracking. Notably, the sharp rise in maximum velocity error at $300\,\text{kg}$ reflects the effects of inertia and actuator power saturation, whereas the moderate increase in RMS errors indicates stable overall performance across the payload range.

\section{Conclusion}
\label{sec:conclusion}
This paper presented a comprehensive framework for electrifying and achieving sensorless control of HDRMs using EMLAs. An integrated EMLA model combining PMSM electromechanics and directional screw-thread efficiency was embedded within a VDC-based joint-space architecture. Actuator sizing was formulated as a multi-objective NSGA-II problem over motor choice, gearbox ratio, and screw lead, accelerated via a deep neural-network model for sub-millisecond actuator efficiency predictions. For sensorless control, a physics-informed Kriging surrogate reconstructed load-side force, velocity, and three-phase currents from only torque and angular velocity measurements. Integration of this surrogate into the VDC controller was experimentally validated on a one-degree-of-freedom testbed, demonstrating accurate trajectory tracking under varying loads. The results confirm that the Kriging observer eliminates the need for dedicated force and velocity sensors, while the VDC control ensures precise performance across a wide operating range. By uniting detailed modeling, multi-objective design, and sensorless control, this work lays the foundation for replacing hydraulic actuation with efficient, zero-emission EMLAs in HDRMs. 


\bibliography{References}

\bibliographystyle{IEEEtran}

\vspace{-0.5cm}
\begin{IEEEbiography}[{\includegraphics[width=1in,height=1.25in,clip,keepaspectratio]{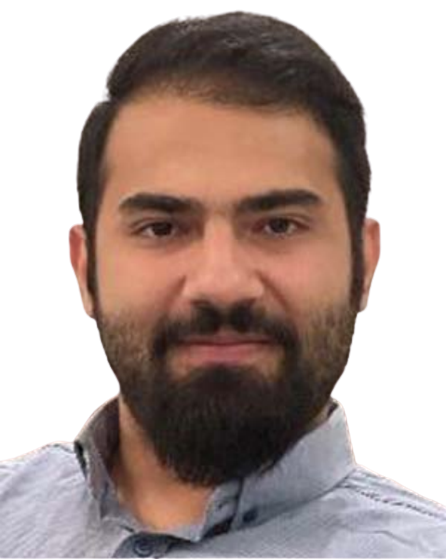}}]{Mohammad Bahari} earned a B.Sc. in electrical engineering power from Semnan University, Semnan, Iran, in 2015, followed by the completion of his M.Sc.  in electrical engineering power electronics and electric machines from Sharif University of Technology, Tehran, Iran, in 2019. Presently, he is engaged as a doctoral researcher at Tampere University, Tampere, Finland, focusing on design and control of an all-electric robotic e-boom. His research interests include multidisciplinary design optimization of electromechanical actuator.  
\end{IEEEbiography}
\vspace{-0.5cm}

\begin{IEEEbiography}[{\includegraphics[width=1in,height=1.25in,clip,keepaspectratio]{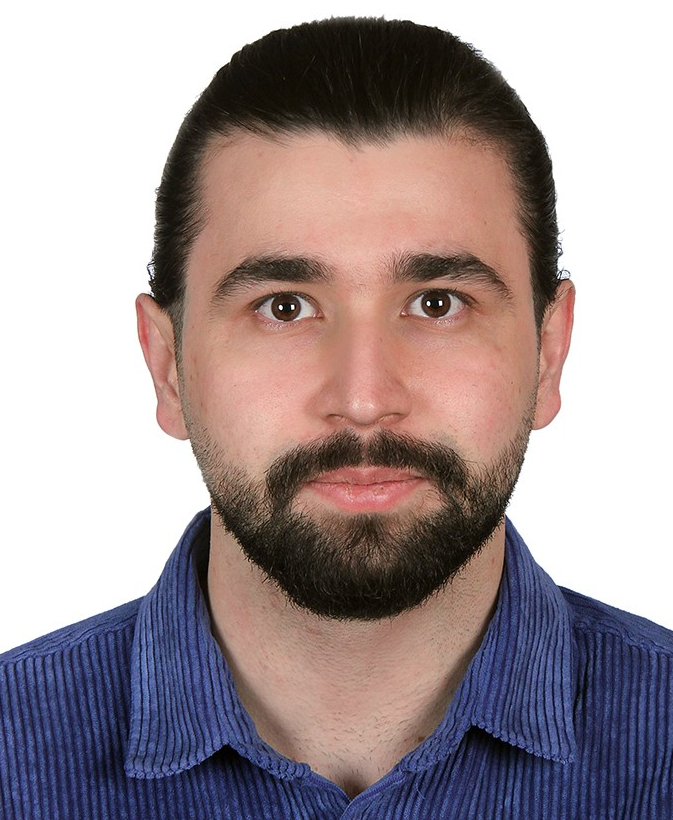}}]{Amir Hossein Barjini} received a B.Sc. in Mechanical Engineering from Amirkabir University of Technology (AUT), Tehran, Iran, in 2021, and an M.Sc. in Mechanical Engineering from Sharif University of Technology (SUT), Tehran, Iran, in 2024. Currently, he is a doctoral researcher at Tampere University, Tampere, Finland. His research interests include applications of artificial intelligence in control, automation, and robotics.
\end{IEEEbiography}
\vspace{-0.5cm}

\begin{IEEEbiography}[{\includegraphics[width=1in,height=1.25in,clip,keepaspectratio]{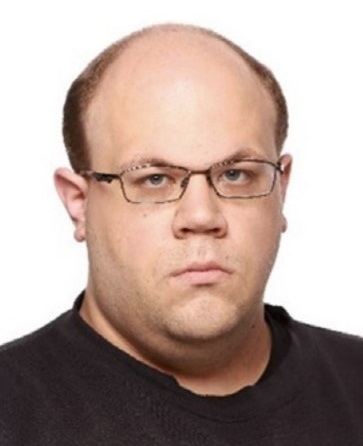}}]{Pauli Mustalahti} received his M.Sc. degree in engineering from the Tampere University of Technology in 2016 and his D.Sc. (Tech.) degree in Automation Science and Engineering from Tampere University in 2023. He is a researcher in the Unit of Automation Technology and Mechanical Engineering, Tampere University, Tampere, Finland. His research interests include nonlinear model-based control of robotic manipulators.
\end{IEEEbiography}
\vspace{-0.5cm}

\begin{IEEEbiography}[{\includegraphics[width=1in,height=1.25in,clip,keepaspectratio]{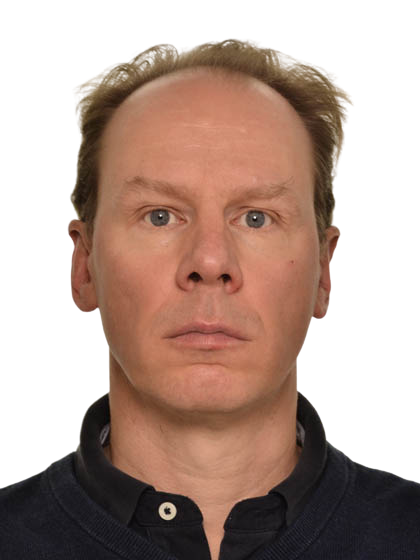}}]{Jouni Mattila}
received an M.Sc. and Ph.D. in automation engineering from Tampere University of Technology, Tampere, Finland, in 1995 and 2000, respectively. He is currently a professor of machine automation with the Unit of Automation Technology and Mechanical Engineering at Tampere University. His research interests include machine automation, nonlinear-model-based control of robotic manipulators, and energy-efficient control of heavy-duty mobile manipulators.
\end{IEEEbiography}

\vfill

\end{document}